\documentclass[WileyMSP-template,citeautoscript,reprint,floatfix]{revtex4-2}
\usepackage{comment}
\usepackage[]{graphicx}
\usepackage{dcolumn}%
\usepackage{bm}%
\usepackage{float}%
\usepackage[utf8]{inputenc}

\let\baraccent=\= 
\renewcommand{\=}[1]{\stackrel{#1}{=}} 

\begin{document}
\setcitestyle{super}

\title{Defect Density of States of Tin Oxide and Copper Oxide \textit{p}-type Thin-film Transistors}

\author{ Måns J. Mattsson$^1$, Kham M. Niang$^2$, Jared Parker$^1$, David J. Meeth$^2$, John F. Wager$^3$, Andrew J. Flewitt$^2$ and Matt W. Graham$^{1*}$}
\affiliation{$^{1.}$Department of Physics, Oregon State University, Corvallis, OR, USA} 
\affiliation{$^{2.}$Electrical Engineering Division, University of Cambridge, Cambridge, UK} 
 \affiliation{$^{3.}$School of EECS, Oregon State University, Corvallis, OR, USA}
 \affiliation{$^*$E-mail: graham@physics.oregonstate.edu }
\begin{abstract}
\indent	
\\
The complete subgap defect density of states (DoS) is measured using the ultrabroadband (0.15 to 3.5 eV) photoconduction response from p-type thin-film transistors (TFTs) of tin oxide, SnO, and copper oxide, Cu$_2$O.  The TFT photoconduction spectra clearly resolve all bandgaps that further show the presence of interfacial and oxidized minority phases.  In tin oxide, the SnO majority phase has a small 0.68 eV bandgap enabling ambipolar or p-mode TFT operation.  By contrast, in copper oxide TFTs, an oxidized minority phase with a 1.4 eV bandgap corresponding to CuO greatly reduces the channel hole mobility at the charge accumulation region. Three distinct subgap DoS peaks are resolved for the copper oxide TFT and are best ascribed to copper vacancies, oxygen-on-copper antisites, and oxygen interstitials. For tin oxide TFTs, five subgap DoS peaks are observed and are similarly linked to tin vacancies, oxygen vacancies, and oxygen interstitials. Unipolar p-type TFT is achieved in tin oxide only when the conduction band-edge defect density peak ascribed to oxygen interstitials is large enough to suppress any n-mode conduction. Near the valence band edge in both active channel materials, the metal vacancy peak densities determine the hole concentrations, which further simulate the observed TFT threshold voltages.

 
\end{abstract}
\keywords{tin oxide; copper oxide; CMOS; thin-film transistor; subgap density of states; photoconduction}

 \maketitle  
\section{Introduction}

   Reliable p-type metal oxide materials remain elusive despite widespread adoption of n-type metal oxide thin-film transistors (TFTs) for applications such as display panels.  Several fundamental and practical barriers prevent p-type TFTs from achieving the same high mobility and low leakage currents that are now celebrated in n-type materials such as amorphous indium gallium zinc oxide (a-IGZO).\cite{nathan2014amorphous,wakimura2015simulation, wager2014amorphous,hosono2018we} Most metal oxide semiconductors are inherently oxygen-deficient and thus n-type.  Even at large gate voltages, p-type behavior is difficult to achieve due to the deleterious effects of very large Urbach energies ($E_U=$ 90-120 meV), with correspondingly large hole trap densities.  To show how subgap defect states enable successful p-mode operation in metal oxide TFTs, this work measures the subgap integrated trap density using the ultrabroadband photoconductive density of states (UP-DoS) method.\cite{vogt2020ultrabroadband,mattson_Adv}
  
  Over the past decade, many p-type metal oxide semiconductors have been suggested, including NiO, TeO$_2$, CuO, and different types of spinel oxides.\cite{wang2016recent,ouyang2022research,devabharathi2024alpha,xu2019p} Two of the most promising active channel material candidates for p-mode oxide TFTs are tin and copper oxide with a typical hole mobility of 1-2 $\text{cm}^2\,\text{V}^{-1}\,\text{s}^{-1}$ and reported on-to-off ratios of $\sim 10^3-10^5$. \cite{lee2023cu2o,fortunato2010transparent,ogo2008p}  While considerable improvement of p-type TFTs is anticipated,  p-type TFT device properties are not yet commensurate with typical n-mode metal oxide TFTs that now have mobilities  $>$10 $\text{cm}^2\,\text{V}^{-1}\,\text{s}^{-1}$ and on-to-off ratios approaching  $10^{12}$.\cite{lan2011high,nomura2004room}  The discovery of more viable metal oxide p-type TFTs remains critical for beyond-silicon complementary metal oxide semiconductor (CMOS) needed for conventional von Neumann computing architectures. Likewise, emerging neuromorphic computer architectures and memory technologies seek to exploit the metal and oxygen vacancy defects in p-type TFTs.\cite{jang2022amorphous,belmonte2021tailoring}

  \par One necessary condition for robust p-mode TFT operation is sharp valence band tail states with small Urbach energies. In metal oxides, such small tail energies at valence band maximum are rare owing to an overwhelmingly large anionic disorder created by oxygen 2p-orbital interactions. In addition to the large effective mass of the asymmetric 2p-orbitals limiting mobility, the resulting broad valence band maximum Urbach tails lead to a high concentration of localized tail states, which act as hole traps. \cite{iordanidou2021optoelectronic,wager2017real}
  Unusual for metal oxides, orbital theory in tin and copper oxide predicts that the valence band maximum band-edge states may have strong Sn-5s and Cu-3d orbital character that may compete with oxygen-2p disorder to form the highest occupied molecular orbital.\cite{hu2022interlayer,wang2016electronic} The diffuse and symmetric orbitals of the Sn-5s and Cu-3d contributions further help explain the higher mobilities and smaller hole effective mass reported in the literature.\cite{fortunato2010transparent,fortunato2010thin, hu2022interlayer,sekkat2021open}

Electronic structure and density functional theory (DFT) simulations widely attribute metal vacancies as the origin of preferential effective p-type doping of tin oxide and copper oxide TFTs.\cite{togo2006first, lee2020hydrogen,allen2013understanding,vzivkovic2020exploring,scanlon2009acceptor}  There is considerable disagreement between different DFT studies on the exact structure and metal vacancy placement within the subgap relative to the many other significant defect peaks predicted.\cite{togo2006first, lee2020hydrogen,allen2013understanding,vzivkovic2020exploring,scanlon2009acceptor}  Previous experimental studies use field-effect conductance or CV-methods, and provide valuable insight on the near band regions only.\cite{lee2020switching,jeong2015subgap,luo2015control}  Instead, the UP-DoS method uses scanning laser microscopy over a tunable $h\nu$=0.15 to 3.5 eV range to obtain a photoconduction (PC) spectrum proportional to the total integrated trap density, and thereby obtain the subgap defect density of states (DoS) for tin and copper oxide TFTs.   
\par 
  \begin{figure}
\begin{center}
   \begin{tabular}{c}
   \includegraphics[width=3.3 in.]{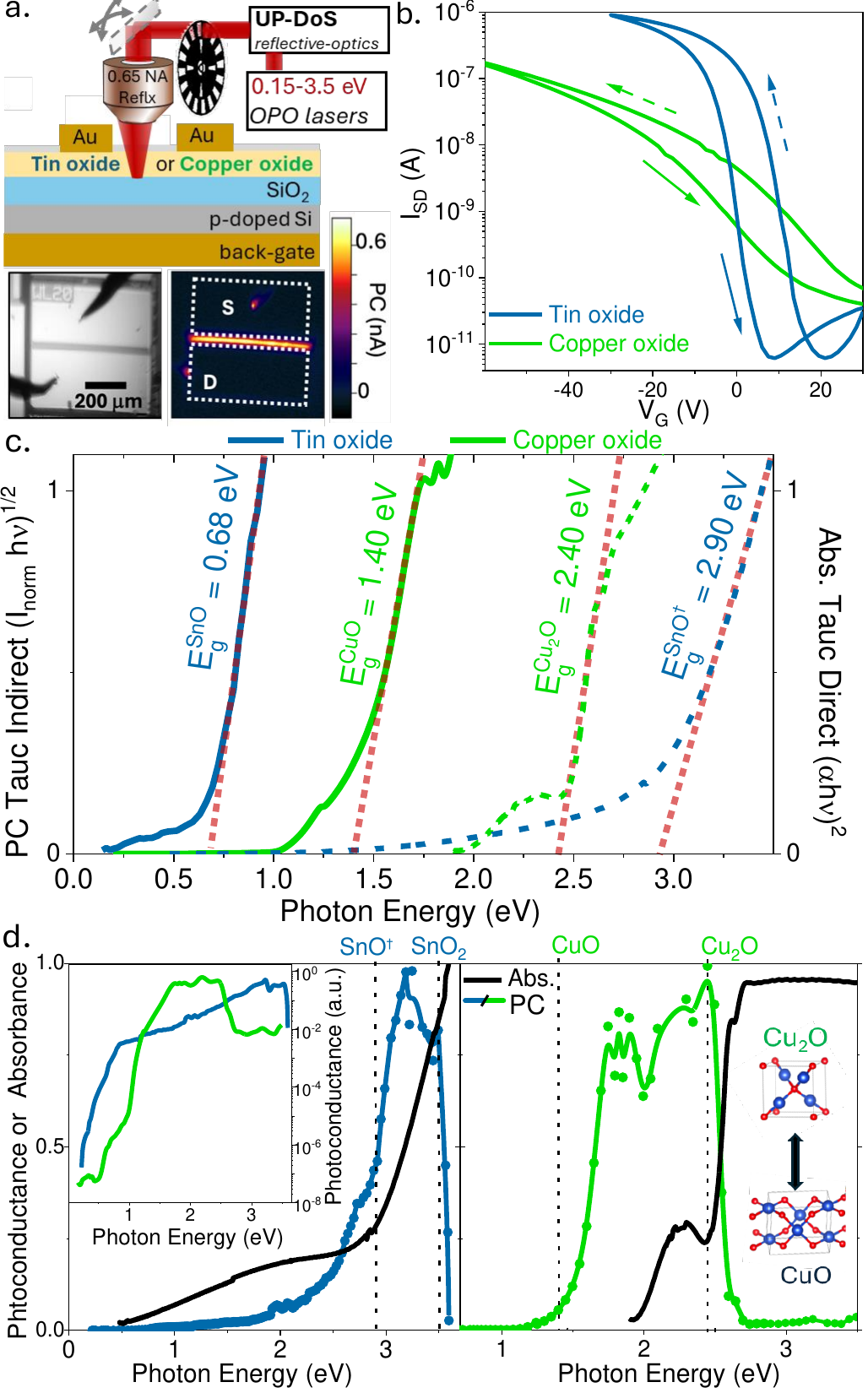}
   \end{tabular}
   \end{center}
     \caption{\textbf{(a)} (\textit{top}) TFT layer cross section sketched with the basic UP-DoS measurement setup shown. (\textit{bottom}) Copper oxide TFT reflection and photoconductance (PC) spatial maps for $h\nu$ = 1.1 eV.  \textbf{(b)} TFT transfer curves of tin (\textit{blue}) and copper (\textit{green}) oxide. \textbf{(c)}  Solid lines show normalized TFT PC plotted with Tauc-like scaling [$(I_{\text{norm}}h\nu)^{1/2}$] yielding an indirect bandgap of 0.68 eV corresponding to SnO and 1.4 eV corresponding to CuO. Dashed lines plot the Tauc absorption spectra scaled as [($\alpha h\nu)^{1/2}$] revealing a direct bandgap of 2.9 eV corresponding to SnO, and 2.4 eV corresponding to Cu$_2$O. \textbf{(d)} At the labeled SnO$_2$ and Cu$_2$O direct bandgaps, the absorption spectrum (\textit{black}) rises, while the PC-spectrum (\textit{blue/green}) falls. (\textit{inset}) The same TFT PC-spectra plotted on a logarithmic scale.  
      } 
\end{figure}
\begin{center}
\begin{table}
 \begin{tabular}{c c c  c } 
 & Absorbance & Photoconduction  & Literature\\  & $E_g^{\text{abs}}$ (eV) & $E_g^{\text{PC}}$ (eV) & $E_g$ (eV) \\

 \hline

 SnO & - & 0.68 & 0.70-0.75 \cite{wu2022exploiting,saji2016p} \\

SnO$^{\text{\textdagger}}$ &  2.9 & 2.9 &2.8-3.0 \cite{guo2010microstructure}  \\

SnO$_2$$^{\text{\textdagger}}$  & - & 3.5 & 3.5-3.7 \cite{saji2016p,zhou2014band}\\


CuO & - & 1.4 & 1.5 \cite{khoo2020electronic}\\

Cu$_2$O$^{\text{\textdagger}}$ & 2.4 & 2.4 & 2.0-2.6 \cite{ozaslan2020effect,khoo2020electronic}\\
\hline

 \end{tabular}
  \caption[example]
 { \label{bandgap} Measured and literature bandgaps for tin and copper oxides. $^{\text{\textdagger}}$ Direct-bandgap.}

 \end{table}
\end{center}
\section{Results and Discussion}
Figure 1a shows the cross section of tin and copper oxide TFTs with their associated photoconduction (PC) microscopy map in the lower panel. The measured transfer curves (W/L = 10, V$_{SD}$ = 1 V) shown in Fig. 1b of tin oxide and copper oxide TFTs both show characteristic p-mode behavior with a negative gate voltage turn-on and $\sim$10$^5$ and $\sim$10$^4$ on-to-off ratios, respectively.  Clockwise hysteresis due to hole trapping and re-emission is pronounced for both types of TFTs. 
Figure 1c presents indirect Tauc-like plots of normalized TFT photoconductance [$(I_{\text{norm}}h\nu)^{1/2}$ versus h$\nu$] that are shown as solid lines. For tin oxide, the estimated indirect bandgap is equal to 0.68 eV, corresponding to SnO, in good agreement with reported values for SnO of 0.70-0.75 eV. \cite{wu2022exploiting,saji2016p}  For copper oxide, the estimated indirect bandgap is found to be 1.4 eV, corresponding to CuO. Direct Tauc plots [($\alpha h\nu)^{1/2}$ versus h$\nu$] are also shown in Fig. 2a and are indicated as dashed lines. A direct bandgap of 2.9 eV is found for tin oxide, corresponding to SnO, while a direct bandgap of 2.4 eV is estimated for copper oxide, consistent with Cu$_2$O. Table \ref{bandgap} summarizes the bandgaps retrieved from photoconduction and absorption measurements, along with their comparisons of DFT calculated values reported in the literature.
\begin{figure*}[!htb]
    \centering
    \includegraphics[width=1 \linewidth]{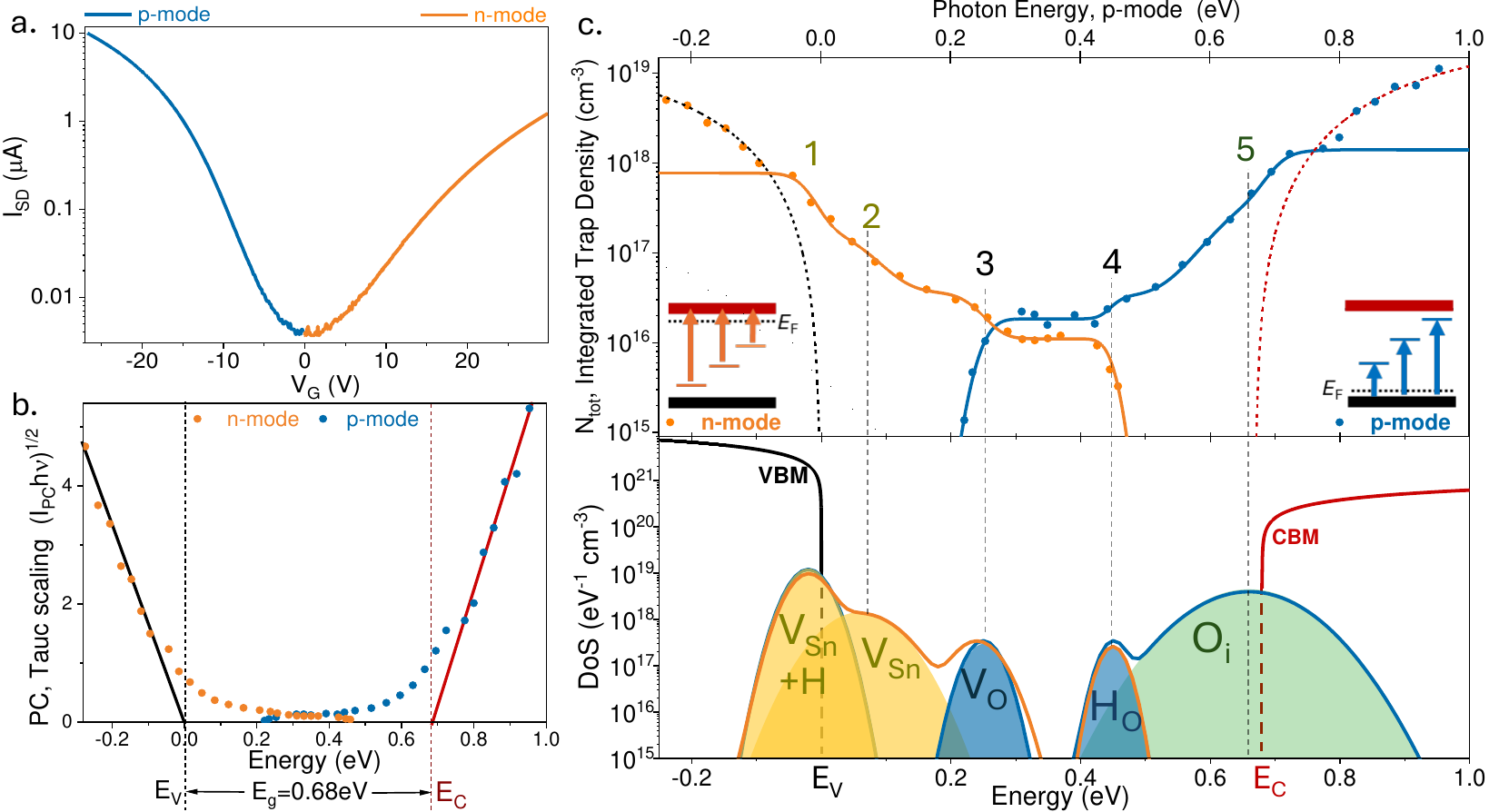}
 \caption{ \textbf{Ambipolar tin oxide subgap DoS} \textbf{(a)} TFT transfer curve of ambipolar SnO (W/L = 100) at V$_{SD}$ = 1 V for p-mode (blue) and n-mode (orange) operation. \textbf{(b)}  Tauc scaling of the photoconduction taken in n-mode (\textit{orange}) and p-mode (\textit{blue}) showing an SnO bandgap of 0.68 eV. \textbf{(c)} \textit{Upper panel} plots the SnO integrated trap density, N$_{tot}$, measured at p-mode and n-mode TFT gate voltages. \textit{Lower panel} plots the corresponding DoS of SnO that reveals five numbered subgap peaks with suggested defect assignments according to Table \ref{SnO}. 
}
    \label{fig:ambi}
\end{figure*}
\begin{center}
\begin{table*}
 \begin{tabular}{c c c  c c c c} 
DoS & Assigned & Peak Energy & FWHM  & Peak DoS  &   DFT Formation &   DFT Peak  \\
 Peak & Defect & (eV) & (meV)  &$\mathrm{\times 10^{17}}$$\mathrm{cm^{-3} {eV^{-1}}}$&  Energy\cite{varley2013ambipolar} (eV) &  Energy\cite{varley2013ambipolar} (eV) \\ 
 \hline

1 &  V$_{\text{Sn}}$+H& $\mathrm{-0.02}$ & $\mathrm{25}$ & $\mathrm{120.6}$ & $\mathrm{0.6}$ & $\mathrm{0.07}$\\

2 & V$_{\text{Sn}}$ & $\mathrm{0.06}$ & $\mathrm{46}$ & $\mathrm{13.3}$ & $\mathrm{1.5}$  & $\mathrm{0.13}$ \\

3 & V$_{{\text{O}}}$ & $\mathrm{0.25}$ & $\mathrm{21}$ & $\mathrm{3.2}$ & $\mathrm{1.4}$ & $\mathrm{0.24}$ \\

4 & H$_{O}$& $\mathrm{0.45}$ & $\mathrm{16}$ & $\mathrm{2.7}$ & $\mathrm{1.2}$  & $\mathrm{0.60}$ \\

5 &   O$_{i}$ & 0.66 & 
63 & 39.2 & 1.1  & $>0.7$\\
\hline
 \end{tabular}
  \caption[example]
 { \label{SnO} 
 Estimated subgap DoS parameters from UP-DoS measurements of the ambipolar SnO TFT plotted in Fig. 2c are compared to  DFT calculated values by Varley \textit{et al}. \cite{varley2013ambipolar}
}
 \end{table*}
\end{center}
\par
Figure 1d shows a comparison of absorption (\textit{black}) and photoconductance (\textit{blue} for tin oxide; \textit{green} for copper oxide). Abrupt thresholds in which absorption increases while photoconductance decreases of 3.5 eV for tin oxide and 2.4 eV for copper oxide provide strong evidence for the presence of SnO$_2$ and Cu$_2$O, respectively. The \textit{inset} of  Fig. 1d  further shows the total raw photoconductance spectrum drops off by a factor of 10$^5$ for tin oxide and 10$^7$ for copper oxide, as subgap defect states are photoexcited below the respective indirect bandgaps of SnO and CuO.
\par
Taken together, Figs. 1c and 1d provide strong evidence that both the tin oxide and copper oxide thin-films are mixed-phase systems corresponding to SnO $\&$ SnO$_2$ and CuO $\&$ Cu$_2$O, respectively. Since XRD characterization shows no evidence of SnO$_2$ and CuO in these thin-films, we conclude that CuO and SnO$_2$ are minority phases compared to Cu$_2$O and SnO.\cite{gomersall2023multi,han2016effects} For copper oxide, the measured Hall mobility ($\mu_{\text{Hall}}=$ 20 cm$^2$/Vs) is consistent with Cu$_2$O while the measured field-effect mobility ($\mu_{\text{FE}}=$ 0.1 cm$^2$/Vs) is most likely due to CuO. Similar mobility properties have been widely reported in the literature. \cite{jo2020causes,guo2015influences,singh2023cvd} As TFT photoconductance is primarily sensitive to the charge accumulation layer, this supports that the Cu$_2$O-SiO$_2$ (semiconductor-dielectric) interface is heavily oxidized to CuO. \cite{im2014photo} \cite{ran2015analyses}
\subsection{Ambipolar SnO Subgap DoS   }

Figure \ref{fig:ambi}a plots the ambipolar SnO TFT transfer curve, and delineates the p-mode and n-mode conduction region where the UP-DoS spectra will be taken.  The SnO bandgap is identified in Fig. \ref{fig:ambi}b by applying Tauc scaling to the raw $I_{\text{norm}}(h\nu)$ signal.  The extracted bandgap of $E_g=0.68$ eV is in good agreement with previously reported 0.70-0.75 eV SnO bandgaps.\cite{wu2022exploiting,saji2016p}  Figure \ref{fig:ambi}c (\textit{upper panel}) plots the integrated trap density, N$_{tot}(h\nu)$, for p-mode (blue line, V$_G=-20$ V) and n-mode (orange line, V$_G=27$ V) operation.  N$_{tot}(h\nu)$ is directly proportional to measured TFT photoconductance, and the subgap region consists of a sequence of step-wise  features, where each step corresponds to a Gaussian peak in the subgap DoS shown by the dashed lines connected to the \textit{lower panel}. Each subgap step in N$_{tot}(h\nu)$ is analytically fit to an error (erf) function (solid line). The above-bandgap response is fit to an indirect absorption lineshape function (dashed lines) accounting for valence band to conduction band transitions. Note that plotting both the p-mode and n-mode signal on the same energy axis requires that the n-mode energy axis be shifted by $E_V+E_{h\nu}=-(E_C-E_{h\nu}-E_g)$, where $E_V+E_{h\nu}$ (VBM = 0) is the energy-axis of p-mode, and $E_C-E_{h\nu}$ (CBM = 0) is the energy axis of n-mode. 
\par
Figure \ref{fig:ambi}c (\textit{lower panel}) plots the SnO subgap DoS across the full bandgap. This is quite extraordinary since all prior UP-DoS measurements were conducted using unipolar semiconductors, where measuring the critical subgap states within a few k$_b$T to the active band edge was not possible.
Note that the DoS shown in the \textit{lower panel} of Figure 2c is obtained by differentiating \textit{upper panel} data, i.e., DoS =  $\frac{dN_{tot}}{d(h\nu)}$. Since each step in the N$_{tot}$ data is fit to an erf, each N$_{tot}$ step yields a corresponding Gaussian in the DoS plot. The use of error function fitting prior to differentiating significantly improves the DoS signal-to-noise ratio compared to simple point-by-point numerical differentiation of the raw N$_{tot}$ data.
Table \ref{SnO} summarizes the parameters of each subgap DoS peak, numbered 1-5. 

 \par
The identities of SnO subgap peaks 1-5 in Fig. 2c are ascribed by comparison to the DFT calculations of Varley \textit{et al.}.\cite{varley2013ambipolar} Subgap peaks in SnO are assigned as: 1. tin vacancy + hydrogen defect (V$_{\text{Sn}}+\text{H}$, acceptor), 2. tin vacancy (V$_{\text{Sn}}$, acceptor), 3. oxygen vacancy (V$_{\text{O}}$, donor), 4. hydrogen on an oxygen site (H$_{\text{O}}$, donor) and 5. oxygen interstitial (O$_{i}$, acceptor). While the tin and oxygen vacancy peaks are expected, peaks 1 and 5 require more discussion.
\par
Figure 2c shows peak 1 is observed in both n- and p-mode UP-DoS measurements at a photon energy $0.02$ eV above the bandgap. Thus, in principle, peak 1 could be positioned $0.02$ eV below E$_V$ or above E$_C$. Since peak 1 is observed to have the largest density of all peaks, it is assigned to the defect with the lowest formation energy, V$_{Sn}$+H located at the VB edge, as identified by Varley \textit{et al}.
\par
Varley \textit{et al.} calculates a hydrogen interstitial donor defect with transition energy of $\sim 0.6$ eV from the valence band maximum,\cite{varley2013ambipolar} which could potentially correspond to peak 5 measured at $0.66$ eV from the valence band maximum. 
However, from charge balance considerations, peak 5 needs to be a neutral acceptor rather than a positively charged ionized donor. Therefore, we propose an oxygen interstitial O$_i$ acceptor as an alternative identification for this peak. 
While DFT studies calculate a low formation energy for this defect, they also predict a transition energy above the bandgap rather than near the conduction band minimum. \cite{lee2020hydrogen,varley2013ambipolar,togo2006first,oba2010native} In part owing to the coplanar structure of the SnO lattice layers,  oxygen interstitial are suspected to be abundant in SnO, and function as an intermediary species in the oxidation of SnO to SnO$_2$. \cite{oba2010native,januar2024unveiling}
\par
\begin{figure}
    \centering
    \includegraphics[width=3.4 in.]{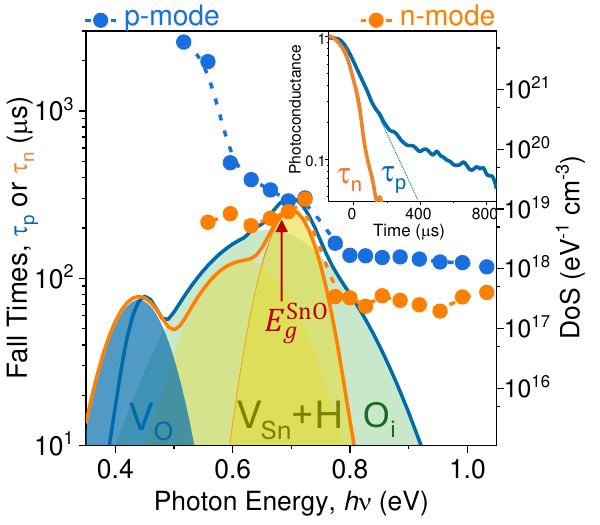}
    \caption{Tin oxide TFT fall times of photoconduction response for photon energies in n-mode operation (\textit{orange}) and p-mode operation (\textit{blue}).  Overlaid are the subgap DoS peaks (right axis) seen in n- and p-mode excitations that show the fall times changes with each successive subgap DoS peak. (\textit{inset}) p- and n-mode time traces of the photoconduction fall after excitation at $h\nu$ = 1 eV and extraction of the dominant exponential fall-times, $\tau_{p/n}$. }

    \label{fig:lifetime}
\end{figure}

 Figure \ref{fig:lifetime} shows TFT fall times, $\tau$ changes for each successive peak in the overlaid SnO DoS until plateauing for above gap photoexcitations at $\tau_p \approx$130 $\mu$s in p-mode (\textit{blue line}) and $\tau_n \approx \sim$70$  \mu$s in n-mode (\textit{orange line}).  Interestingly, near $h\nu \approx0.7$ eV, both n- and p-mode fall times merge to match at $\tau_{p/n} \approx$ 280 $\mu$s, suggesting both photoexcitions are dominated by relaxation from the same band-edge defect site. This matching abrupt increase in both fall times in Fig. \ref{fig:lifetime}, independently confirms the strong carrier trapping nature of the corresponding V$_{\text{Sn}}$+H acceptor DoS band-edge peak that is overlaid from Fig. 2c. Exploring midgap photoexcitations down to 0.5 eV, the extracted SnO TFT fall time $\tau_p$ increases tenfold to $\sim$2.5 ms at p-mode gate voltages. To help explain, previous UP-DoS fall time studies by Vogt \textit{et al}. show charge-neutral defect states have inherently longer fall times than ionized defects from weaker coulombic interactions.\cite{vogt2020ultrabroadband} Accordingly, such p-mode mid-gap photoexcitations are consistent with empty deep donors being photoexcited similar to V$_{\text{O}}^+ + h \nu \rightarrow \text{V}_{\text{O}}^0+h^+$ to create a neutral filled donor with the expectedly slower $\sim$2.5 ms recombination time observed.  
  
\subsection{p-type SnO Subgap DoS}
\begin{figure*}[!htb]
    \centering
    \includegraphics[width=1.0 \linewidth]{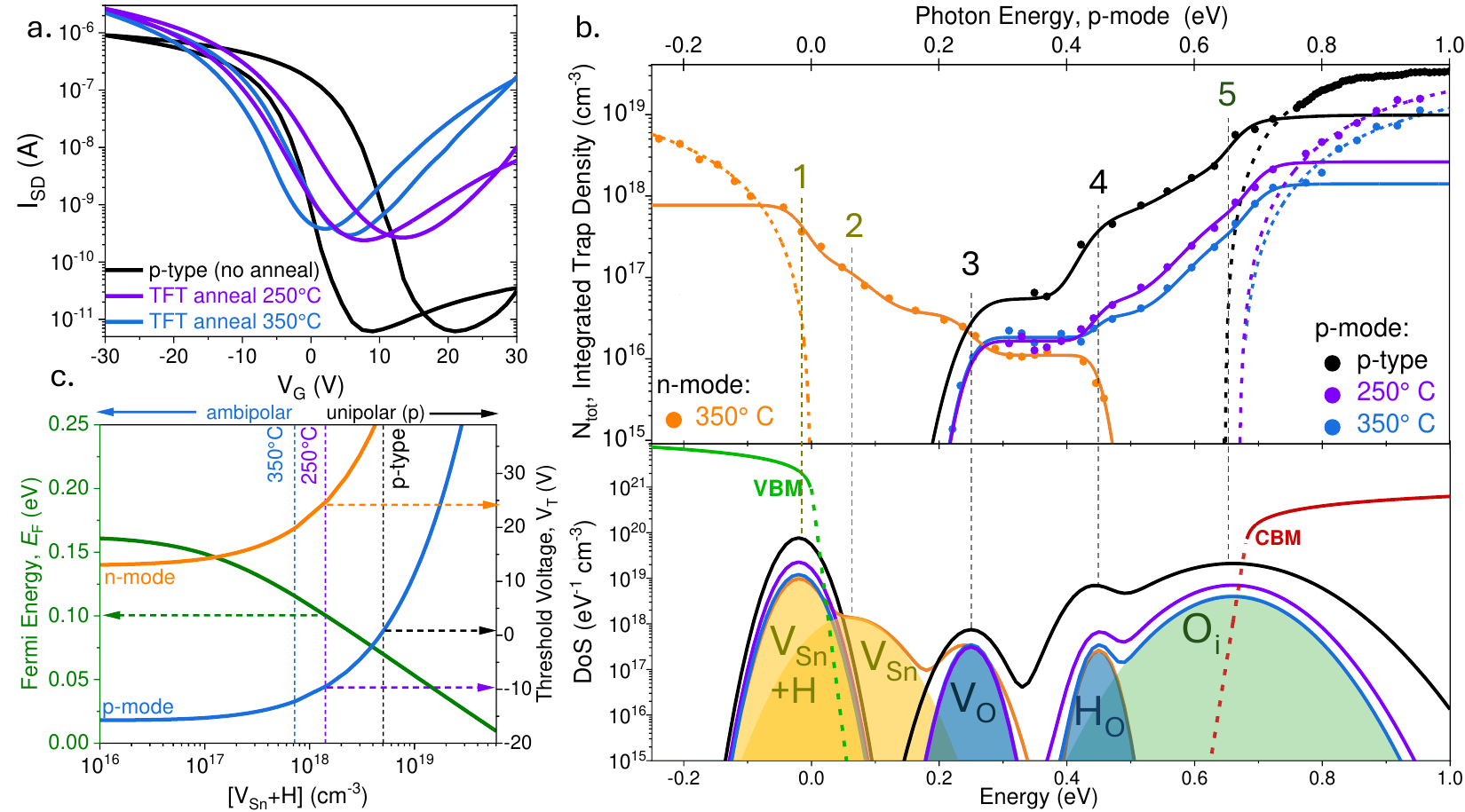}
    \caption{  \textbf{p-type SnO Subgap DoS} \textbf{(a)} SnO TFT (W/L = 10, $V_{SD}$ = 1 V) transfer curves show unipolar p-type TFT behavior with no post-deposition anneal (\textit{black}), and increasing ambipolar TFT behavior after annealing at 250 $^{\circ}$C (\textit{purple}) and 350 $^{\circ}$C (\textit{blue}).    \textbf{(b)}  Experimental integrated trap density N$_{tot}$  (\textit{upper panel}) and subgap DoS (\textit{lower panel}). \textbf{(c)} As the concentration of peak 1, [V$_{Sn}$+H], increases, the simulated equilibrium Fermi level energy, E$_F$ (green), moves towards the valence band and the threshold voltage (blue and orange) shifts positively for both p-mode and n-mode behavior, respectively.
  } 
    \label{fig:p-type}
\end{figure*}
Figure \ref{fig:p-type} shows the evolution of the SnO subgap DoS from p-mode to ambipolar TFT behavior.  In Fig. \ref{fig:p-type}a three SnO TFT transfer curves (W/L = 10, $V_{SD}$ = 1 V) are classified as p-type (black), weak-ambipolar (250 $^{\circ}$C annealed, purple), and strong-ambipolar (350 $^{\circ}$C annealed, blue).  The corresponding integrated trap densities, N$_{tot}$, are shown in Fig. \ref{fig:p-type}b (\textit{upper panel}). Using orange circles, just one example of a n-mode N$_{tot}$ spectrum is plotted in Fig. \ref{fig:p-type}b (\textit{upper panel}) corresponding to the most clearly ambipolar TFT in  \ref{fig:p-type}a (blue) . Indicated by the vertical dashed lines, Fig. \ref{fig:p-type}b (\textit{upper panel}) shows five subgap steps labeled 1–5  that correspond to Gaussian DoS peaks shown in Fig. \ref{fig:p-type}b (\textit{lower panel}). The peaks observed and their identifications match those previously discussed in Fig. \ref{fig:ambi}. 
\par
The \textit{lower panel} of Fig. \ref{fig:p-type}b further shows that as SnO changes with annealing conditions from p-type to ambipolar behavior, a corresponding decrease in subgap defect density is observed. The significant enhancement in n-mode behavior with annealing that is observed in Fig. \ref{fig:p-type}a is attributed primarily to a decrease of the defect peak located just below the conduction band minimum, identified as O$_i$. Note that for n-mode operation, O$_i$ behaves as an electron trap.\cite{wager2022amorphous} To achieve unipolar p-type SnO TFT operation with low off current, similar to black curve of \ref{fig:p-type}a, [O$_i$] should be sufficiently large to suppress the undesired n-mode behavior.
\par
Figure \ref{fig:p-type}c presents a simulation of the Fermi level energy together with the p- and n-mode threshold voltages as a function of [V$_{Sn}$+H]. The dashed vertical lines indicate the measured concentration of each TFT, whereas the horizontal dashed arrows are visual aids to guide each simulated parameter to the correct y-axis. When $[V_{Sn}+H] \gtrsim 10^{17}$cm$^{-3}$ (as observed for all three TFTs), the simulation shows that [V$_{Sn}$+H] controls the Fermi energy, and consequently the threshold voltage, whereas to a good approximation $p \approx [V_{Sn}+H$]. Furthermore, Fig. \ref{fig:p-type}c shows that at $[V_{Sn}+H] \lesssim 10^{18}cm^{-3}$, both n- and p-mode threshold voltages are easily accessible, enabling ambipolar TFT behavior. 
\begin{figure*}
    \centering
    \includegraphics[width=1 \linewidth]{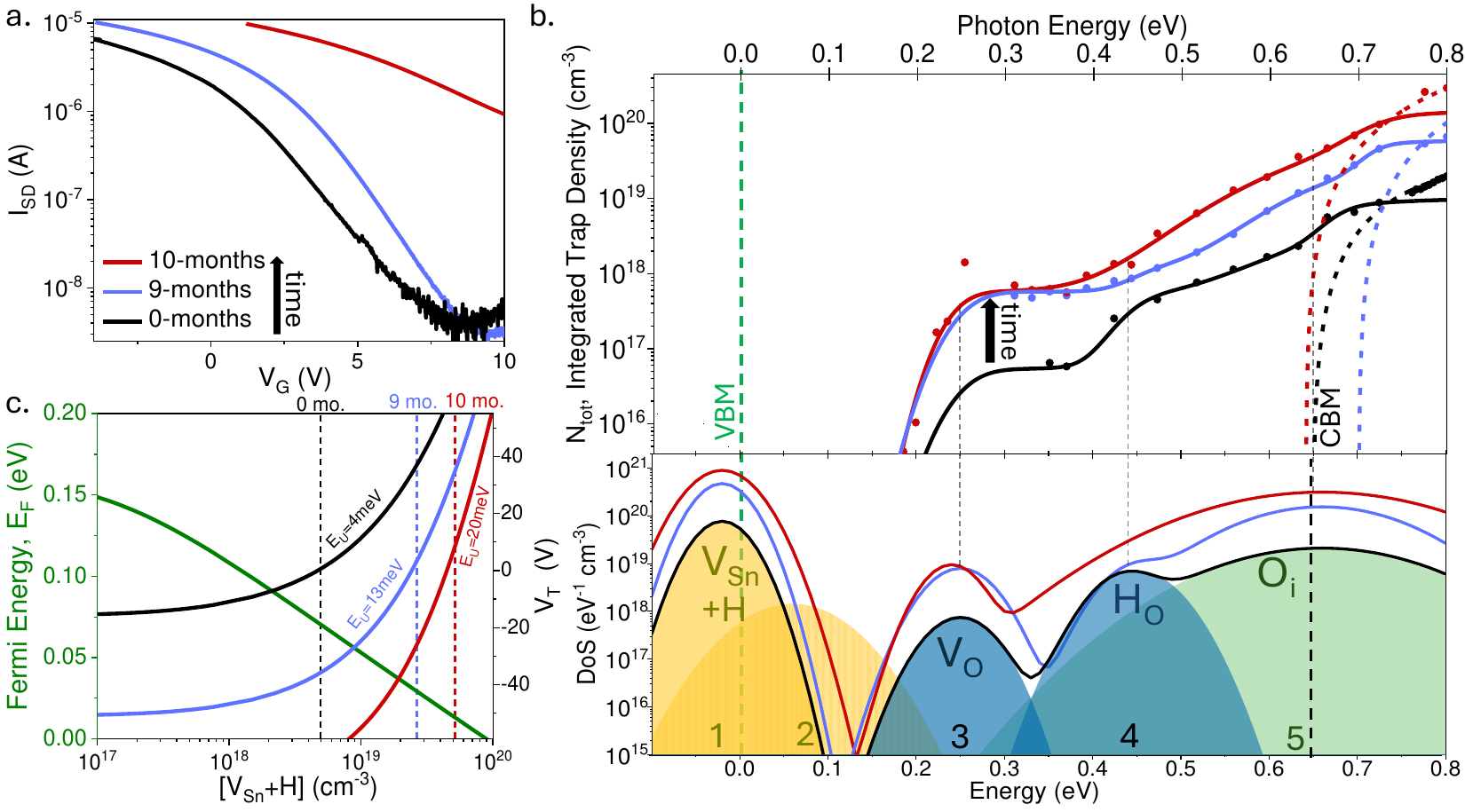}
    \caption{  \textbf{SnO Subgap Trap Density Oxidation.} \textbf{(a)} SnO TFT  ($V_{SD}$ = 1 V, W/L = 100) transfer curve shift with time, from 0-months (black), 9-months (green) to 10-months (cyan) in vacuum storage  \textbf{(b)} The integrated trap density N$_{tot}$ \textit{(upper panel)} measured by UP-DoS and corresponding subgap DoS \textit{(lower panel)} at three vacuum storage times. \textbf{(c)} Simulated Fermi level energy, E$_F$, and threshold voltage, $V_{T}$, reveals that the valence band Urbach energy, E$_U$, increases with time.} 
\label{time}
\end{figure*}
\subsection{ SnO TFT DoS Stability Study }
SnO TFT properties change after 0-, 9-, and 10-months of vacuum storage, with a 1-week exposure to lab atmosphere introduced between months 9 and 10. As indicated in Fig. \ref{time}a, the threshold voltage exhibits a large positive shift from 1 V (black), to 3 V (blue), and finally to 8 V (red). Furthermore, the off current further increased dramatically from $\sim$0.01 $\mu$A at 9-months to $\sim1$ $\mu$A at 10-months. Similarly, with increasing time Fig. 5b shows a substantial increase in the measured N$_{tot}$ (\textit{upper panel}) with a corresponding increase of subgap DoS (\textit{lower panel}). After 10 months, the resulting large increase in the subgap DoS near Peaks 1 suggests the TFT has become so strongly p-type doped that turn-off is unfeasible, as seen in the associated transfer red curve shown in Fig. 5a.
\par

Figure \ref{time}c plots the simulated Fermi level energy and p-mode threshold voltage as a function of increasing concentration of tin-vacancy related states, [V$_{\text{Sn}}$+H]. Previously, in the Fig. \ref{fig:p-type}c simulation, the valence band Urbach energy extracted was constant at E$_U \sim 4$ meV over the three different TFT processing conditions. However, the extracted Urbach energy increases in Fig. \ref{time}c with the subgap density from E$_U \sim 4$ meV at 0 months to E$_U \sim 13$ meV at 9 months, and finally to E$_U \sim 20$ meV after 10 months of vacuum storage.

Since the valence band Urbach energy is a measure of disorder on the anion sublattice, it is reasonable that there may be some correlation between certain subgap peaks and valence band tail states. The observed increase in [V$_{\text{Sn}}$+H] with increasing vacuum storage time is likely associated with the oxidative process,  SnO $\rightarrow$ SnO$_{2}$ since this process is known to be associated with tin vacancy formation. \cite{leijtens2017mechanism,hao2014lead}  However, it is also possible that hydrogen incorporation or diffusion could also play a role in SnO TFT degradation. 
\begin{figure}
    \includegraphics[width=1 \linewidth]{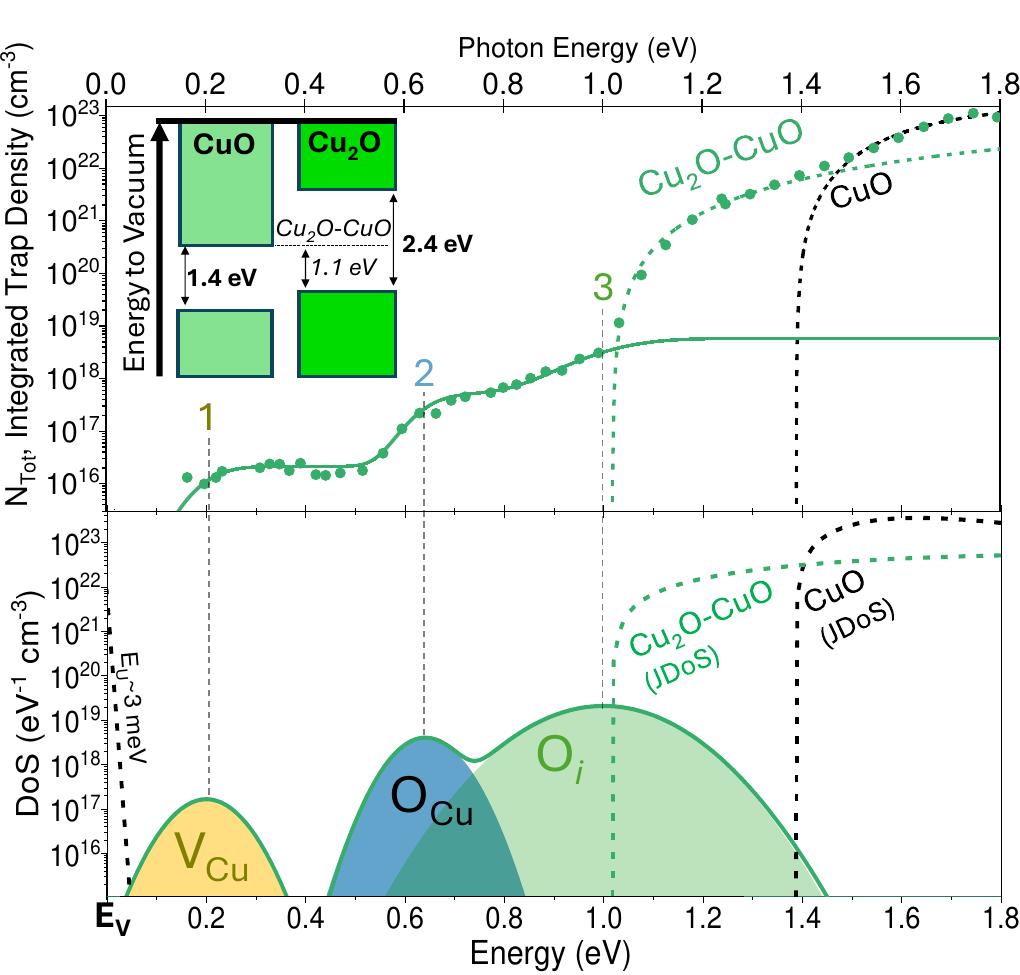}
    \caption{  Copper oxide TFT integrated trap density, N$_{tot}$, (\textit{upper panel}) and density of states, DoS, (\textit{lower panel}).  The above-bandgap UP-DoS response is fit to indirect transition lineshapes (\textit{dashed}), while the below-bandgap response displays three clear steps that are fit to error functions. The DoS plot reveals three subgap peaks with suggested defect identifications, as summarized in Table III. Band alignment diagram for the mixed-phase system is shown as an \textit{inset}.}
    \label{fig:cuodos}
\end{figure}
\subsection{Copper Oxide Subgap DoS}
The copper oxide  TFT transfer curve is unipolar p-type with a 10$^4$ turn-on, as shown previously in Fig. 1b.  Figure \ref{fig:cuodos} presents the associated subgap integrated trap density, N$_{tot}$ in the top panel, and density of states, DoS in the bottom panel.   Three N$_{tot}$ clear steps labeled 1-3 are indicated in Fig. \ref{fig:cuodos} along with three corresponding Gaussian DoS peaks positioned at 0.20, 0.64, and 1.00 eV above the valence band maximum. Proposed defect identifications for these three peaks are labeled in Fig. \ref{fig:cuodos} and summarized in Table III.
\par
Proposed peak identifications of the three copper oxide peaks measured are based on DFT calculations by Zivkovi{\'c} and de Leeuw, or by Scanlon \textit{et al.}, for CuO or Cu$_2$O defect site and formation energies. \cite{vzivkovic2020exploring, scanlon2009acceptor}  The measured 0.20 eV peak is identified as a copper vacancy acceptor, V$_{\text{Cu}}$, comparable to the 0.17 eV value estimate for CuO or the 0.23 eV estimate for Cu$_2$O. The measured 0.64 eV peak is ascribed to an oxygen-on-copper antisite defect, O$_{\text{Cu}}$, similar to the calculated value of 0.49 eV on the CuO sublattice. \cite{vzivkovic2020exploring}  Finally, the peak measured at 1.00 eV is ascribed to an oxygen interstitial, O$_i$, since both phases are calculated to have O$_i$ point defects in this portion of the bandgap with predicted ionization energies of 1.14 eV and 1.08 eV for CuO and Cu$_2$O, respectively.\cite{vzivkovic2020exploring, scanlon2009acceptor} Note that the O$_i$ defect in copper oxide, as shown in Fig. \ref{fig:cuodos}, has a similar Gaussian width and position within the subgap as the O$_i$ defect in tin oxide, as shown in Fig. \ref{fig:ambi}.
\par
Similar to tin oxide, the measured copper oxide subgap DoS is used to estimate the Fermi level energy and valence band Urbach energy to be E$_F$ = 260 meV and E$_U$ = 3 meV, respectively. Then, from charge balance, the copper oxide equilibrium hole concentration is estimated to be $p =2 \times10^{16} \text{cm}^{-3}$, which is approximately equal to the copper vacancy concentration $p\approx$ [V$_{Cu}$].
\begin{center}
\begin{table*}
 \begin{tabular}{c c c  c c c c} 
DoS & Assigned & Peak Energy & FWHM  & Peak DoS  &   DFT Formation & DFT Peak \\
 Peak & Defect & (eV)   & (meV)  &$\mathrm{\times 10^{17}}$$\mathrm{cm^{-3} {eV^{-1}}}$&  Energy\cite{vzivkovic2020exploring} (eV) &  Energy\cite{vzivkovic2020exploring} (eV) \\

 \hline

1 &  V$_{\text{Cu}}$& $\mathrm{0.20}$ & $\mathrm{50}$ & $\mathrm{1.6}$ & $\mathrm{1.0}$ & $\mathrm{0.17}$ \\

2 & O$_{\text{Cu}} $ & $\mathrm{0.64}$ & $\mathrm{48}$ & $\mathrm{40}$ & $\mathrm{0.5}$ & $\mathrm{0.49 }$\\

3 & O$_{i}$ & $\mathrm{1.00}$ & $\mathrm{99}$ & 210 & 0.3  & $\mathrm{1.14}$ \\

 \hline
 \end{tabular}
  \caption[example]
 { \label{CuO} 
Estimated subgap DoS peak parameters from UP-DoS measurements of the copper oxide TFT plotted in Fig. 6c. Experimental results are compared to DFT calculated values by Zivkovi{\'c} and de Leeuw \cite{vzivkovic2020exploring}
}
 \end{table*}
\end{center}
Recall from Fig. 1c that the indirect bandgap of copper oxide was estimated to be 1.4 eV – equal to that of CuO – based on a linear-regression fit to a Tauc-like plot of normalized photoconductance. However, note that this indirect bandgap estimate is based on ignoring a weaker subgap absorption response with a lower-energy threshold of $\sim$ 1.1 eV (see Fig. 2a). If this weaker subgap absorption response is taken into account, as indicated in Fig. \ref{fig:cuodos}, it can only be included as an above-bandgap feature (in which the DoS is given by a quadratic function of energy), rather than a below-bandgap feature (in which the DoS is described by a Gaussian function of energy). This suggests that the indirect bandgap of copper oxide is actually equal to 1.1 eV rather than 1.4 eV, as expected for CuO. 
\par
How do we rationalize the physics of this 1.1 eV lower-energy threshold since it appears to be inconsistent with both CuO and Cu$_2$O? We propose that this 1.1 eV lower-energy threshold is a mixed-phase transition involving real-space electron transfer in which a filled initial electronic state at the Cu$_2$O valence band maximum is excited into an empty final electronic state at the CuO conduction band minimum. Such a process is accomplished by real-space electron transfer from the Cu$_2$O (filled) initial state into the CuO (empty) final state, and is labeled in Fig. \ref{fig:cuodos} as a Cu$_2$O-CuO (mixed phase) transition. Khoo \textit{et al.} estimate the energy difference from the Cu$_2$O valence band to the CuO conduction band to be 1.09 eV, in good agreement with our observed 1.1 eV threshold energy. \cite{khoo2020electronic} Figure \ref{fig:cuodos} \textit{inset} shows the estimated energy band alignment for the mixed-phase copper oxide.
\begin{figure}[H]
    \includegraphics[width=1 \linewidth]{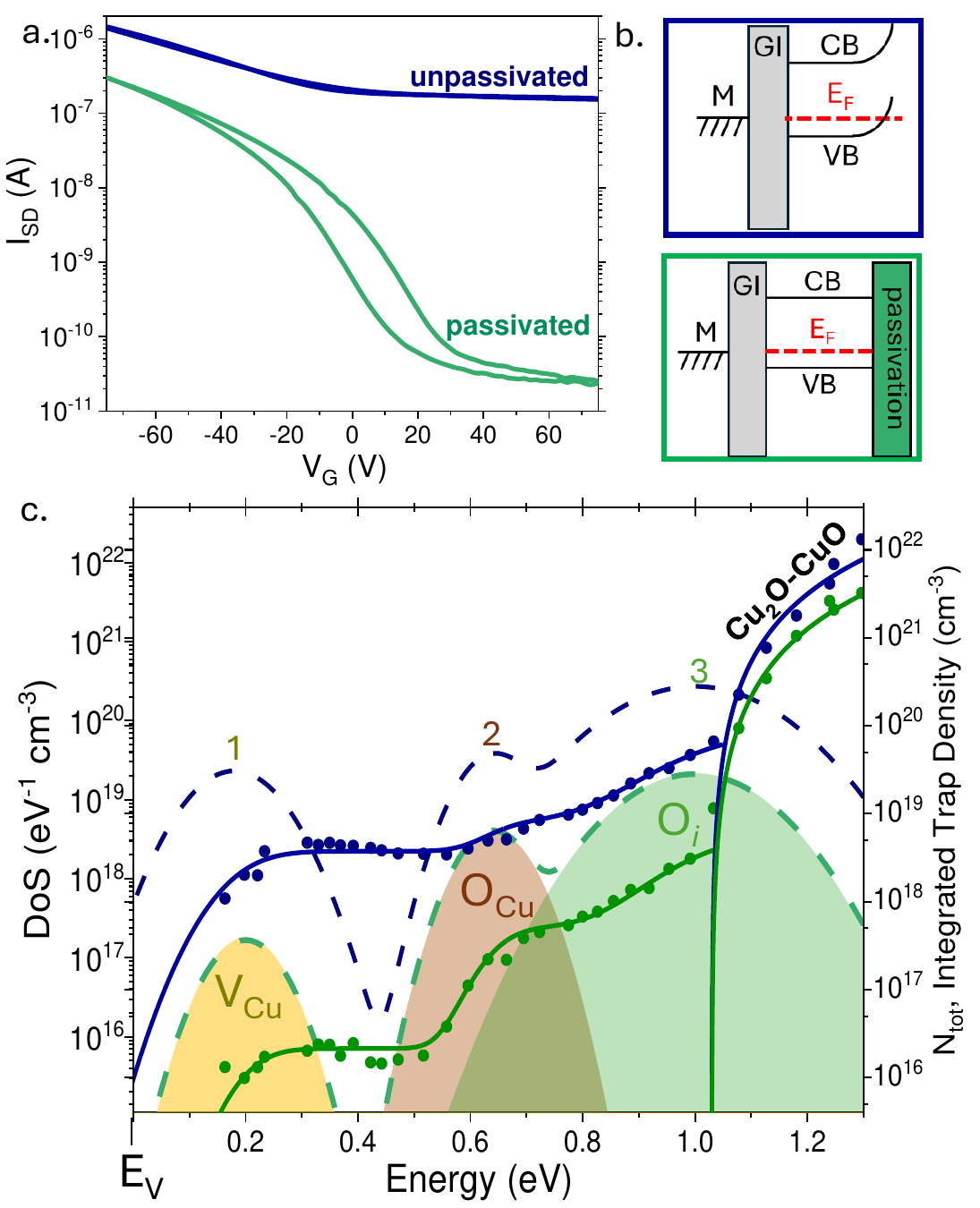}
    \caption{ \textbf{(a)} Transfer curves of unpassivated and passivated copper oxide TFTs at $V_{SD} = 1$ V (W/L = 10). \textbf{(b)} Band diagrams illustrate the strong back-channel band bending expected in the unpassivated device at equilibrium. \textbf{(c)} DoS (dashed lines, left axis) and integrated trap density (solid lines, right axis) of the unpassivated and passivated copper oxide TFT.
     }
     \label{unpassivated}
 \end{figure}
\subsection{Unpassivated and Passivated Copper Oxide Subgap DoS }
Figure \ref{unpassivated}a compares the transfer curves of copper oxide TFTs when the back-channel is passivated (\textit{green}) or unpassivated (\textit{blue}). 
The unpassivated device has a high off current and no clear turn-off voltage.
Figure \ref{unpassivated}b illustrates energy band diagrams showing the existence of a backside accumulation layer responsible for the poor performance exhibited by the unpassivated TFT. In contrast, passivation eliminates the back-side accumulation layer and its deleterious effect upon TFT performance. Figure \ref{unpassivated}c presents a comparison of N$_{tot}$ and DoS for an unpassivated and passivated copper oxide TFT. Consistent with prior studies, the degraded performance of the unpassivated TFT appears to correlate with the enhanced defect DoS across the entire subgap. \cite{napari2021role}  This result suggests that an increase in charged copper vacancies at the backside may be responsible for the formation of this additional accumulation layer.

\section{Conclusions}

\par

Tin oxide and copper oxide constitute two possible material strategies for realizing p-type oxide TFTs.  For each TFT, the subgap integrated trap density is measured over a tunable $h\nu=$ 0.15 - 3.5 eV laser excitation range using the ultrabroadband photoconduction DoS (UP-DoS) technique.\cite{vogt2020ultrabroadband,mattson_hydrogen_2022,mattson_Adv}  The below-gap photoconductivity drops off by $\sim$10$^5$ in tin oxide and $\sim$10$^7$ in copper oxide and indicates how the corresponding SnO and CuO material phases drive p-type conductivity in each TFT. 

 Measuring subgap integrated traps in n-mode and p-mode for ambipolar TFTs provides two complimentary DoS spectra that fully span the subgap. In tin oxide, the resulting subgap DoS shows five subgap peaks assigned as: tin vacancies (V$_{\text{Sn}}$ and V$_{\text{Sn}}$+H), oxygen vacancies in the midgap (V$_{\text{O}}$ and H$_{\text{O}}$), and oxygen interstitial sites.  The near VB-edge tin vacancy V$_{\text{Sn}}$+H peak largely determines the TFT equilibrium Fermi level energy and p-doping, such that $p\approx [\text{V}_{\text{Sn}+\text{H}}]$. Adjacent to the CB-edge, the broad oxygen interstitial peak density suppresses n-mode conduction to achieve the desired unipolar p-mode TFT behavior with low off current.

 For p-type copper oxide TFTs, the subgap DoS shows just three subgap peaks assigned as: copper vacancies near the VB-edge, oxygen-on-copper antisites in the mid-gap, and oxygen interstitial sites near the CB-edge.  Similar to the tin oxide case, the metal vacancy peak, V$_{\text{Cu}}$, determines the TFT p-doping, such that $p\approx [V_{\text{Cu}}]$. Copper oxide thin-film formation tends to favor the formation of Cu$_2$O, a wider bandgap p-type semiconductor with some inadvertent incorporation of CuO, a narrow bandgap p-type semiconductor with a poor mobility. Even though the oxidized CuO is likely only prevalent near the semiconductor-dielectric interface, its lower bandgap is likely responsible for the low field-effect mobility in the copper oxide TFTs. Synthesis of a phase-pure Cu$_2$O thin-film appears to be a challenging but desirable goal.

\section{Experimental Section}

\textbf{1. p-type TFT preparation} \\
Bottom gate, top contact TFTs were fabricated with a thermally-grown SiO$_2$ (thickness, $t = 200$ nm) as the gate dielectric and heavily-doped p-type Si (100) substrate as the gate electrode. Thermally evaporated gold ($t =100$ nm) was used as the source and drain contacts. The channel layer and the contacts were patterned by photolithography with either lift-off or wet etching method. Unless otherwise stated, the back channel was passivated with Al$_2$O$_3$ ($t = 20$ nm) by atomic layer deposition (ALD) in all devices. For a width-to-length ratio (W/L) of 10, the TFTs reported have a channel width and length of 1000 $\mu$m and 100 $\mu$m.
\par
The Cu$_2$O channel layer ($t = 260$ nm) was deposited in a high target utilization sputtering system from a metallic target. \cite{han2016effects} A post deposition annealing was performed at 700 $^{\circ}$C after the channel patterning. The unpassivated Cu$_2$O TFT exhibits a field effect mobility, $\mu_{FE}$ of 0.21 cm$^2$V$^{-1}$s$^{-1}$ with a small on-off ratio $\sim 10$. In the passivated device, the on-off ratio increases significantly to $\sim 10^4$ but with a reduced on-state current, thus $\mu_{FE}$ of 0.07 cm$^2V^{-1}s^{-1}$.
\par
The SnO channel layer was deposited by ALD using Sn(II) bis-(tert-butoxide) and water at a substrate temperature of 170 $^{\circ}$C.\cite{gomersall2023multi} By varying SnO thickness and process conditions, TFTs exhibit either depletion mode (for TFTs, D1 and D2) or enhancement modes (E1 and E2), as shown in Supporting Table S.I. TFTs incorporating a thicker SnO layer are expected to have a high carrier concentration (D1 and D2), therefore the device operates in depletion mode and the current conduction (mobility) depends on the annealing temperatures. On the other hand, devices E1 and E2 incorporate a more resistive and thinner SnO layer, thus requiring two annealing steps. Even though the $\mu_{FE}$ is similar, E2 with a higher TFT annealing temperature exhibits ambipolar properties, while E1 with a lower TFT annealing temperature exhibits predominantly p-type characteristics.

Post-growth analysis suggests that Cu$_2$O and SnO should be the dominant materials present in each TFT.\cite{han2016effects,gomersall2023multi}  However, the active channel is known to readily oxidize to CuO and SnO$_2$.\cite{alajlani2017characterisation, kwok2023conversion} Before measurements, the absorption spectra (Cary UV-Vis-IR) for tin and copper oxide thin-films derived from the TFT growths are compared to the corresponding TFT photoconduction(PC) spectra. 
\par
\medskip
\textbf{2. TFT Ultrabroadband Photoconductance DoS Microscopy} \\
The Ultrabroadband Photoconduction Density of States (or UP-DoS) microscopy method is applied to tin and copper oxide TFTs to measure the subgap defects density.\cite{vogt2020ultrabroadband,mattson_hydrogen_2022,mattson_Adv}  Figure 1a depicts the UP-DoS setup measuring the photoconduction, $I_{PC}(h \nu)$ on a TFT with an ultrabroadband tunable laser from photon energies of $h\nu=$ 0.15 to 3.5 eV. The diffraction-limited laser illuminates a tin or copper oxide active channel from the top side through the thin passivation layer of aluminum oxide. The TFT gate voltage is carefully selected to ensure that the TFT transfer curve is a quasi-linear regime, such that the small incident photon flux ($\sim 10^{13}$ photons/cm$^2$) induces a small linear shift in the illuminated transfer curve.\cite{vogt2020ultrabroadband} The TFT conductivity is continuously monitored with a custom-built RF probe microscopy station that uses all-reflective 4\textit{f}-scanning optics. A high-NA (0.45) 52x reflective objective delivers a nearly diffraction-limited tunable-laser. The bottom panel of Figure 1a shows this 1.1 eV laser mapping the TFT $I_{PC}$ response over the copper oxide TFT active region.

To measure the  DoS of a TFT, the primary observable is the photon-normalized photoconduction, $I_{\text{norm}}(h \nu)= h\nu I_{PC}/P$, where I$_{PC}$ is the Zurich lock-in amplifier (HFLI) detected photoconduction signal and $P$ is the incident laser power (ranges from $\sim$ 0.1 to 100 $\mu$W). I$_{\text{norm}}$ has units of AeV$^{-1}$W$^{-1}$.  Providing $h\nu <E_g$, only band-to-defect (p-mode) or defect-to-band (n-mode) transitions are likely, and the detected signal $I_{\text{norm}}$ can be approximated as,
 \begin{eqnarray}
  I_{\text{\text{norm}}}^{\mathrm{p-mode}}(h\nu) \propto N_{filled} ^{VB} \int^{\infty}_{-\infty} N_{empty}(E+h\nu)dE\\
     I_{\text{\text{norm}}}^{\mathrm{n-mode}}(h\nu) \propto N_{empty} ^{CB}\int^{\infty}_{-\infty} N_{filled}(h\nu-E)dE
\end{eqnarray} where $N_{empty}(E+h\nu)$ is the subgap DoS of empty defect states excited to with photon energy $h\nu$ from filled valence band states, $N_{filled} ^{VB}$, in p-mode operation. $N_{filled}(h\nu-E)$ is the subgap DoS of filled defect states that are excited to empty conduction band states, $N_{empty} ^{CB}$, in n-mode operation. See Supporting Information Section I for a comparison when $h\nu>E_g$. 

To obtain the absolute integrated total defect density per unit volume, $N_{tot}$, the directly measure quantity I$_{\text{norm}}(h\nu)$ must be simply rescaled by a constant given in the bracketed expression below,\cite{vogt2020ultrabroadband}
\begin{equation}
    N_{tot}(h \nu)=I_{\text{norm}}(h\nu)\left[ \frac{C_{ox} k_{o}}{qd}\left( \frac{\partial I_{D}}{\partial V_{G}}\right)^{-1} \right]
    \label{eqn:ubpc}
\end{equation}
where $C_{ox}$ is the gate oxide capacitance of the TFT, $q$ is the charge of the electron, $d$ is the thickness of the excited accumulation channel ($\sim$ 0.3  nm). $(\frac{\partial I_{D}}{\partial V_{G}})$ is the non-illuminated slope of the TFT transfer curve in the linear regime where the UP-DoS measurement was conducted.  $k_{o}$ is the rate of incident photons that causes the $I_{PC}$ signal to saturate owing to state filling (or $k_o \approx \frac{P_{sat}}{h\nu}$ ). Finally, the first derivative of Eqn. \ref{eqn:ubpc} with respect to energy provides the desired experimental subgap DoS as $\frac{dN_{Tot}}{d(h\nu)}$.
\par

\textbf{3. Charge Balance and Discrete Trap Model}\\
By using the full functional form of all measured subgap DoS peaks, it is possible to use a discrete trap model together with charge balance to calculate the equilibrium Fermi energy and Urbach energies. The equation for charge balance involves a balance between positive and negative charge,
\begin{equation}
\label{eqn:fermi}
    N^{+}_{GD}(E_F)+N^+_{TD}(E_F)+p(E_F)
    =N^{-}_{GA}(E_F)+n(E_F)
\end{equation}
where ($N^{+}_{GD}$) is ionized Gaussian donors, ($N^{+}_{TD}$) is ionized valence band tail donors, ($N^{-}_{GA}$) is filled Gaussian acceptors, and $n/p$ is free electron/hole density (see Supporting Information Section II for full functional form). The equation for the discrete acceptor trap model is: \cite{hong2008electrical}
\begin{equation}
\label{eqn:model}
  V_{T}(E_F) = -\frac{q}{C_I} \left[N_{TD}^0{}^{\frac{2}{3}} + N_{GD}^0{}^{\frac{2}{3}} + N_{GA}^-{}^{\frac{2}{3}}  - p^{\frac{2}{3}}\right]
\end{equation}
where $q$ is electron charge, and $C_I$ is the gate insulator capacitance density. Note that the only two parameters not measured by UP-DoS in Eq. \ref{eqn:fermi} and Eq. \ref{eqn:model} are the fermi level energy, E$_F$, and the valence band tail donors ($N^{+}_{TD}$ or $N^{0}_{TD}$), whereas ($N^{+}_{TD}$ or $N^{0}_{TD}$) further only depends on E$_F$, the hole effective mass m$_h^*$, and valence band Urbach energy E$_U$. \cite{wager2017real}  By using literature-reported values of m$_h^*$, \cite{ogo2009tin,koffyberg1982photoelectrochemical,hodby1976cyclotron}  the system of equations can be solved for the two unknowns, E$_F$ and E$_U$. Similarly, by instead using the discrete trap donor model,\cite{hong2008electrical} the conduction band Urbach energy can be estimated. Finally, when E$_U$ is known, using the equations above, it is possible to simulate the Fermi level energy together with n- and p-mode threshold voltages as a function of hypothetical change in defect concentrations.

\medskip
\textbf{Supporting Information} \par Details on sample characteristics and optical conductivity data modeling methods.
\medskip

\textbf{Acknowledgments} \par This work is supported by a SAMSUNG Global Research Outreach (GRO) Award. KMN and AJF acknowledge support from EPSRC EP/X025195/1, ‘Innovative Material, Processes, and Devices for Low Power Flexible Electronics: Creating a Sustainable Internet of Everything.’
\medskip

\textbf{Data Availability Statement}: The data that support the findings of this study are available from the corresponding author upon request. 

\medskip


\bibliographystyle{MSP}

\bibliography{mybib.bib}   

\pagebreak

\begin{figure}
\textbf{Table of Contents}\\
\medskip
  \includegraphics{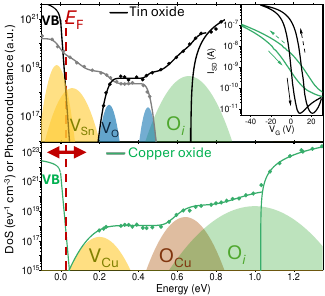}
  \medskip
  \caption{ The complete band-to-band defect density of states driving p-mode operation in tin and copper oxide transistors is obtained through photoconduction microscopy over a 0.15 to 3.5 eV tunable laser range. These analytical maps of subgap defect density with the extracted TFT bandgaps,  show how ideal p-mode operation may be achieved in metal oxide thin-film transistors needed for next-generation CMOS applications.}
  
  
\end{figure}
\end{document}


\title{Supporting Information for: \\ Defect Density of States of Tin Oxide and Copper Oxide \textit{p}-type Thin-film Transistors}
\author{ Måns J. Mattsson$^1$, Kham M. Niang$^2$, Jared Parker$^1$, David J. Meeth$^2$, John F. Wager$^3$, Andrew J. Flewitt$^2$ and Matt W. Graham$^{1*}$}
\affiliation{Department of Physics, Oregon State University, Corvallis, OR 97331-6507, USA} 
\affiliation{Electrical Engineering Division, University of Cambridge, Cambridge CB3 0FA, UK}
\affiliation{School of EECS, Oregon State University, Corvallis, OR 97331-5501, USA}

\maketitle
   \section{Optical Absorption and Photocurrent Response } 

  When $h \nu > E_g$, the I$_{UBPC}$ signal is proportional to the joint density of states (JDoS) as both conduction and valence state densities contribute to the oscillator strength. When the photon energy is smaller than the bandgap in a p-type semiconductor, UP-DoS initial (conducting) states are in the valence band and the final (absorbing) states are subgap defect states (for n-mode operation this is reversed). \begin{figure} [hbt]
   \begin{center}
   \begin{tabular}{c}
   \includegraphics[height=6cm]{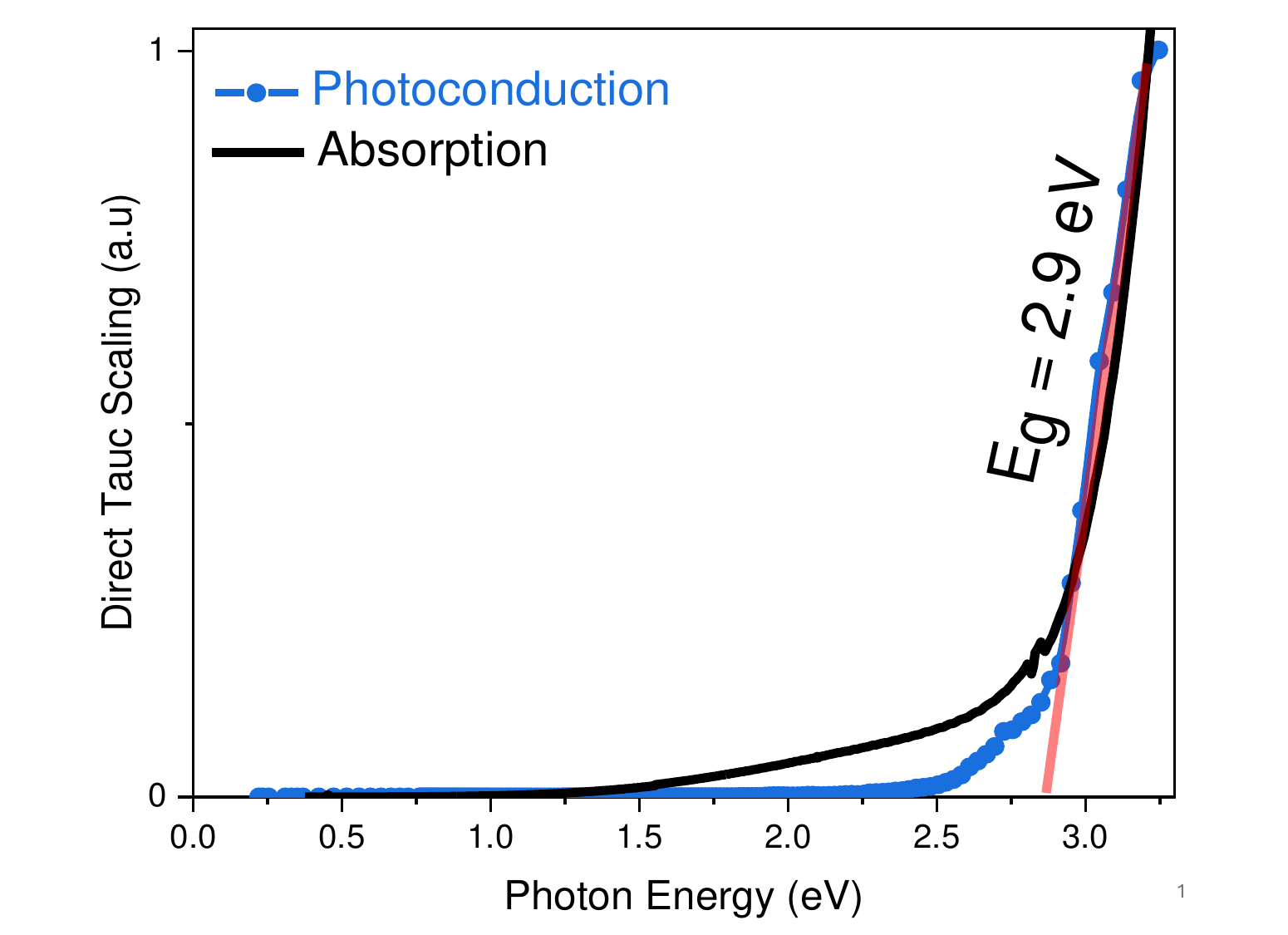}
   \end{tabular}
   \end{center}
   \caption[example] 
   {\textbf{SnO Tauc Analysis} The direct-gap Tauc scaled analysis for SnO shows a bandgap of 2.9 eV for both the TFT photoconduction and optical absorption.} 
   \end{figure} Thus, the subgap UP-DoS signal I$_{UP-DoS}$ is solely established by the final state defect density (or the initial defect state density for n-mode). This can be established quantitatively from the mathematical definition of the absorption coefficient, $\alpha$, as an integral of the initial- and final-state joint density,

   \begin{align}
    \alpha (h \nu) & = \alpha_o \int^{\infty}_{-\infty} N_{filled}^{}(E) N_{empty}(E+h \nu)dE   \\
    &\approx \alpha_o N^{VB}_{filled} \int^{\infty}_{-\infty} N_{empty}(E+h \nu)dE,   \:  \mathrm{ when }  \:  h \nu < E_g  \:  \text{ as here }  \:  N_{empty} \ll N_{filled}  \\
    & \propto I_{UBPC} (h \nu) 
     \end{align} 
 
In UP-DoS the subgap photocurrent response is proportional to the total integrated trap density $N_{traps}$ which means that a different subgap response occurs for p-mode and n-mode TFT operation,
  
  \begin{align}
  I_{UBPC}^{p-mode} (h \nu) \propto N_{traps}(h \nu) \approx N_{filled} ^{VB} \int^{\infty}_{-\infty} N_{empty}(E+h\nu)dE, \:  \text{ for p-mode TFT operation }\\
    I_{UBPC}^{n-mode} (h \nu) \propto N_{traps}(h \nu) \approx N_{empty} ^{CB} \int^{\infty}_{-\infty} N_{filled}(h\nu-E)dE, \:  \text{ for n-mode TFT operation }
     \end{align}

In n-mode operation, optical transitions are from filled defect states near the valence band to empty conduction band states (i.e., defect to CB transitions).  For band-edge analysis using photoconduction, $I_{PC}$, we make the approximation that induced photocurrent is proportional to the photons absorbed in the charge accumulation region, or $I_{PC}\propto N_{h\nu}A$, where $A=1-e^{-\alpha l}$ is the fractional absorption, and $\alpha$ is the absorption coefficient.  This can be expanded as,

 \begin{align}
  I_{PC} (h \nu) & \propto N_{h \nu} \left(1-e^{-\alpha l} \right) \:  \text{ using the absorption expression} \\
  & \approx \alpha N_{h \nu}   l  \: \text{  when $l\ll \alpha$} 
     \end{align}
which is true as the optical skin depth ($\delta = 1/\alpha$) is generally large compared to the thickness of the accumulation region thickness, $l$. In other words, providing $A \ll 1$, then $I_{PC} \propto  \alpha N_{h \nu}$.  Thus, in this limit, standard Tauc and Urbach scaling can be applied to PC-spectral data. 
\par

The above-bandgap region is analogous to a JDoS, and thus is fitted to the textbook direct or indirect optical absorption spectral linshapes. For an indirect bandgap, the absorption coefficient is proportional to \cite{rosencher2002optoelectronics}:
\begin{equation}
    \alpha \propto \frac{(h\nu -E_{\text{g}}+E_{\text{p}})^{2}}{\exp\left(\frac{E_{\text{p}}}{k_bT}\right)-1}+\frac{(h\nu -E_{\text{g}}-E_{\text{p}})^{2}}{1-\exp\left(-\frac{E_{\text{p}}}{kT}\right)}
    \label{indirect}
\end{equation}
where $\alpha$ is the absorption coefficient and $E_{\text{p}}$ is the energy of the accompanying optical phonon. To distinguish optical transitions involving defect states from the band-to-band transitions, the band-to-band photoconduction (for indirect semiconductor) is fit to Eq. \ref{indirect} with some constant of proportionality, while defect involved photoconduction is fit to error functions for each step in N$_{tot}$. The mathematical expression for fitting to a defect step in N$_{tot}$ is:
\begin{equation}
   N_{tot}= A(1+erf[w(E-c)])
\end{equation}
where A is error function amplitude, w is error function width, and c is the error function energy position. Consequently, the derivative of Eq. 9 results in a Gaussian peak in the DoS. Note that the photon energy at which the fitting function transitions from an error function (defect involved transition) to quadratic energy as per Eq. \ref{indirect} (band-to-band transition) results to an unambiguous estimate of the bandgap.
\par
UP-DoS peak assessment is complicated for photon energies near the bandgap since it is not clear whether a near-bandgap subgap peak arises from a near-VBM defect state (initial state) to CB (final state) transition, or a VB (initial state) to near CBM defect state (final state) transition. Considerations beyond UP-DoS data are required to deduce whether a VBM defect $\rightarrow$ CB transition or a VB $\rightarrow$ CBM defect transition is more likely. 
\section{Charge Balance Functional Form}
The equation for balancing positive and negative charge is:
\begin{equation}
\label{eqn:fermi}
    N^{+}_{GD}(E_F)+N^+_{TD}(E_F)+p(E_F)
    =N^{-}_{GA}(E_F)+n(E_F)
\end{equation}
where ($N^{+}_{GD}$) is ionized Gaussian donors, ($N^{+}_{TD}$) is ionized valence band tail donors, ($N^{-}_{GA}$) is filled Gaussian acceptors, and $n/p$ is free electron/hole density. The full functional form of these terms are as follows \cite{wager2017real}:
\begin{equation}
   N^{+}_{GD}(E_F)=n_{GD}\int_{-\infty}^{\infty}{ \exp\left(-\frac{(E - E_{GD})^2}{2w_{GD}^2}\right)(1-f(E))}dE
\end{equation}
\begin{equation}
    N^{+}_{TD}(E_F)=n_{TD} \int_{VME}^{\infty}{ \exp\left(\frac{E_{\text{VME}} - E}{E_U}\right)(1-f(E))}dE
\end{equation}
\begin{equation}
   N^{-}_{GA}(E_F)=n_{GA}\int_{-\infty}^{\infty}{ \exp\left(-\frac{(E - E_{GA})^2}{2w_{GA}^2}\right)f(E)}dE
\end{equation}
\begin{equation}
    p(E_F)=\frac{1}{2 \pi^2} \left( \frac{2m_h^*}{\hbar^2} \right)^{3/2} \int_{-\infty}^{VME}{\sqrt{E_V - E}\:(1-f(E))}dE
\end{equation}
\begin{equation}
    n(E_F)=\frac{1}{2 \pi^2} \left( \frac{2m_e^*}{\hbar^2} \right)^{3/2} \int_{CME}^{\infty}{\sqrt{E - E_C}\:f(E)}dE
\end{equation}
where n$_{GD}$/n$_{GA}$ is Gaussian peak ampltiude, w$_{GD}$/w$_{GA}$ is Gaussian peak width, E$_{GD}$/E$_{GA}$ is Gaussian peak energy, VME/CME is the valence/conduction band mobility edge, E$_V$/E$_C$ is the valence/conduction band energy, m$_h^*$/m$_e^*$ is the hole/electron effective mass, E$_U$ is the valence band Urbach energy, and n$_{TD}$ is the peak density of valence band tail states which is further given by \cite{wager2017real}:
\begin{equation}
    n_{TD}=\frac{1}{2 \pi^2} \left( \frac{2m_h^*}{\hbar^2} \right)^{3/2} \sqrt{\frac{E_U}{2}} 
\end{equation}
Finally, f(E) is the Fermi-Dirac distribution:
\begin{equation}
    f(E) = \frac{1}{1 + \exp\left(\frac{E - E_F}{k_B T}\right)}
\end{equation}
where E$_F$ is the Fermi level energy. Similarly, in the discrete acceptor trap model, the same functional forms as above are used, with the key difference being that filled trap states now matter instead of charged trap states. Specifically, f(E)-1 is replaced by f(E) for all donor states.
\section{Supplementary figure on the UBPC-DoS Experimental Apparatus} 
Continuously scannable laser (mid-IR to the UV) is achieved with multiple femtosecond Ti:Saphire-based lasers.  Specifically, a Coherent Chameleon laser pumps an APE Compact OPO to achieve 680 nm to 4000 nm laser scan range with piezo-mirror optimized Poynting vector stability that keeps the laser-centered on the TFT. To achieve photon energies at the band edge, a home-built difference-frequency generation mixes OPO signal and idler beam overlap on an HGS crystal that extends our range out to 10,000 nm or $h\nu=0.12$ eV  \cite{beutler2016femtosecond}. Lastly, an APE Harmonix unit frequency doubles both pump and signal beams to cover the UV-region of the UBPC microscope's $h \nu$ 0.12 to 3.5 eV continuously scannable range.

\begin{figure}[ht]
   \begin{center}
   \begin{tabular}{c}
   \includegraphics[height=14cm]{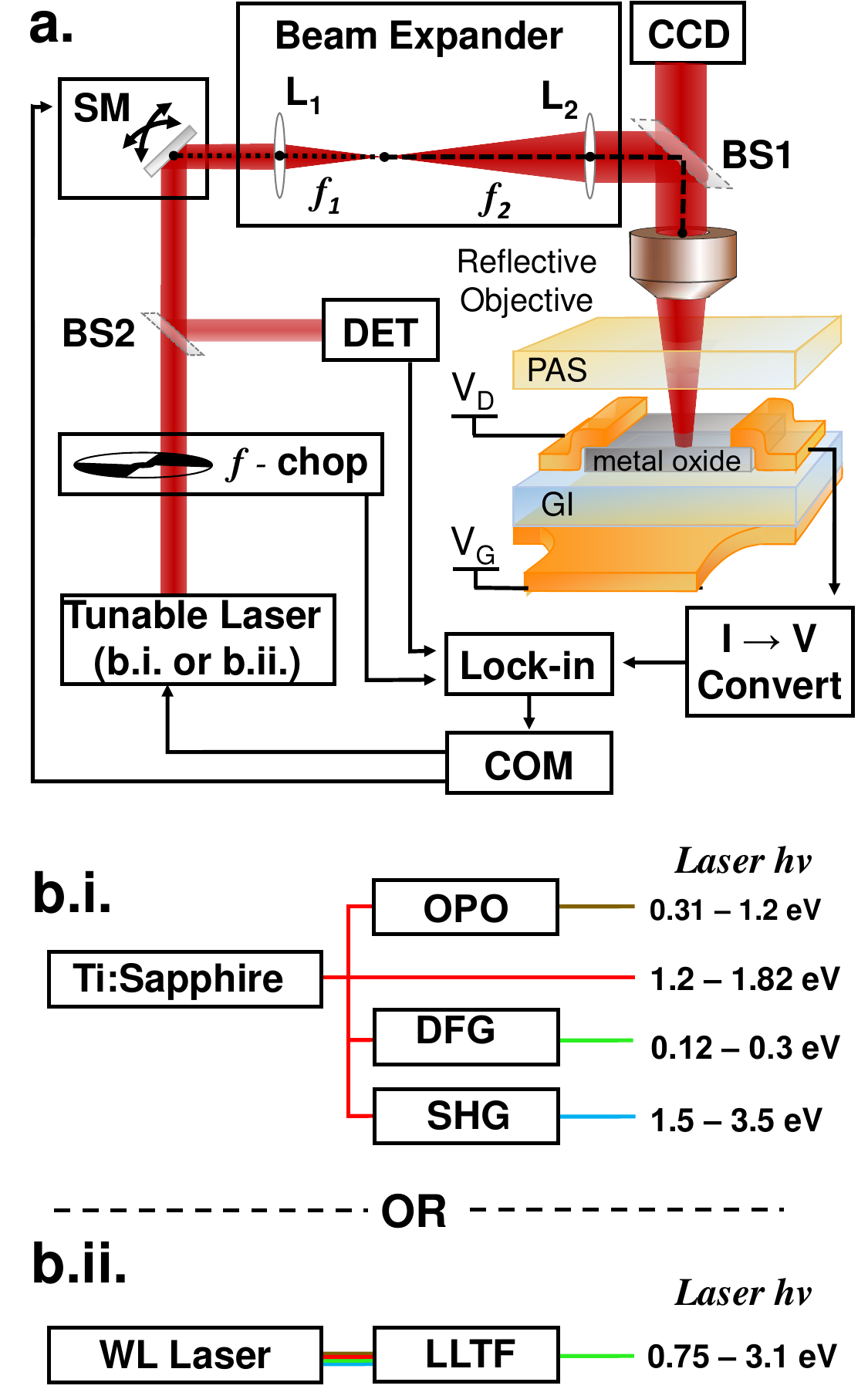}
   \end{tabular}
   \end{center}
   \caption[example] 
   { UBPC Experimental Setup. \textbf{(a)} Tunable lasers coupled into a homebuilt scanning confocal photocurrent microscope allow for continuous photoconduction measurement
over the ranges of the laser excitation sources used. The scanning mirror (SM) and beam-expanding lens pair allow for spatial control of the diffraction-limited laser spot, while a chopper modulates the beam so that the photoconduction can be recorded via lock-in detection. \textbf{(b)} Excitation sources are (i) the Ti:Sapphire-based system, which can cover laser energies from 0.31-3.5 eV, or (ii) the Supercontinuum White-Light (SCWL) laser coupled into a laser line tunable filter (LLTF) which acts as a monochromator.} 
   \end{figure}
   \newpage
\section{Equillbrium Hole Concentration Consistenty Check}
   \begin{figure} [hbt]
   \begin{center}
   \begin{tabular}{c}
   \includegraphics[height=7cm]{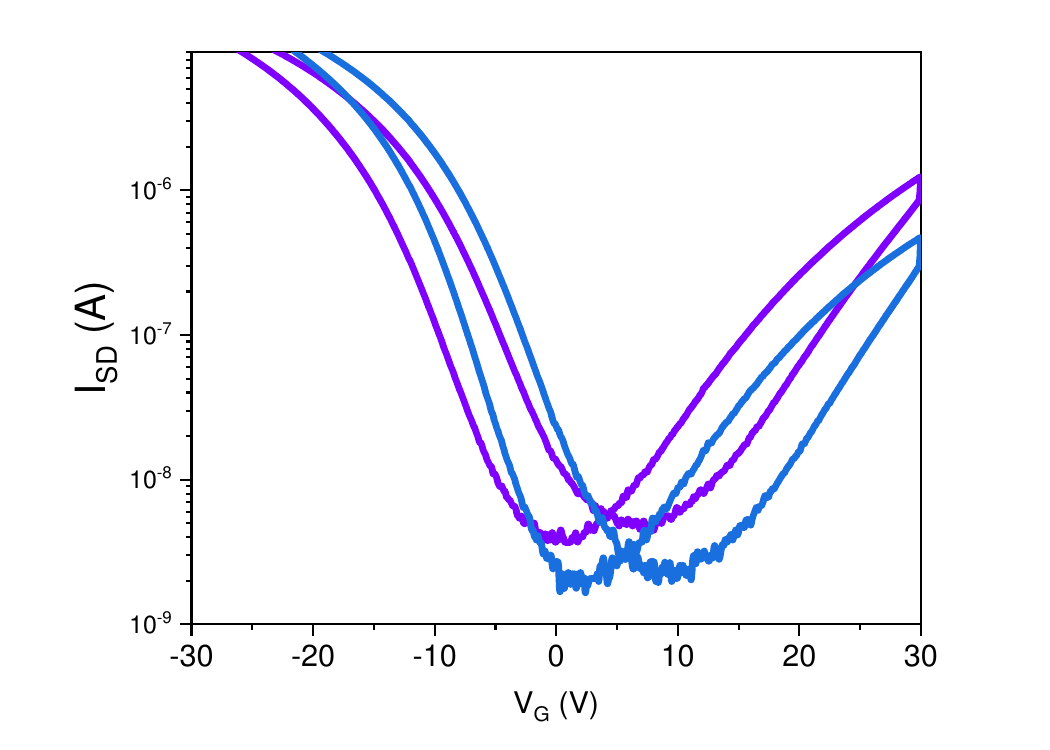}
   \end{tabular}
   \end{center}
   \caption[] 
   {The transfer curves of the E1 (blue) and E2 (purple) were taken at the time of measurement on a W/L=100 device.} 
   \end{figure} 
To independently check how reasonable the $p=0.8$ $\times 10^{18}$ cm$^{-3}$ (E2) and $p=3.6$ $\times 10^{18}$ cm$^{-3}$ (E1) range in free hole density from charge-balance, we can estimate how much $p$ changes between TFTs by comparing the ratio of equilibrium $I_{SD}$ currents.  At the time of measurement, E1 (blue) and E2 (purple) curves have equilibrium source-drain currents of $I_{SD}=2.71$ and $I_{SD}=1.36$ nA on W/L=100 devices.  In agreement, we find, the ratios in charge balance calculated hole concentration, $\frac{p^{blue}}{p^{purple}} \approx 0.54$ and equilibrium $I_{SD}$, $\frac{I^{blue}_{SD}}{I^{purple}_{SD}} \approx 0.50$ are almost identical between SnO TFTs. The unipolar p-type SnO TFT (D1, black) in Fig. 3a also has both the highest equilibrium $I_{SD}$ and calculated hole concentration. However, a similar comparison is not meaningful as the channel thickness of this TFT is also thicker at 30 nm, which also impacts  $I_{SD}$ \cite{li2014effect}.
\newpage
   \section{Saturation Current, $P_{max}$} 
   \begin{figure} [hbt]
   \begin{center}
   \begin{tabular}{c}
   \includegraphics[height=8cm]{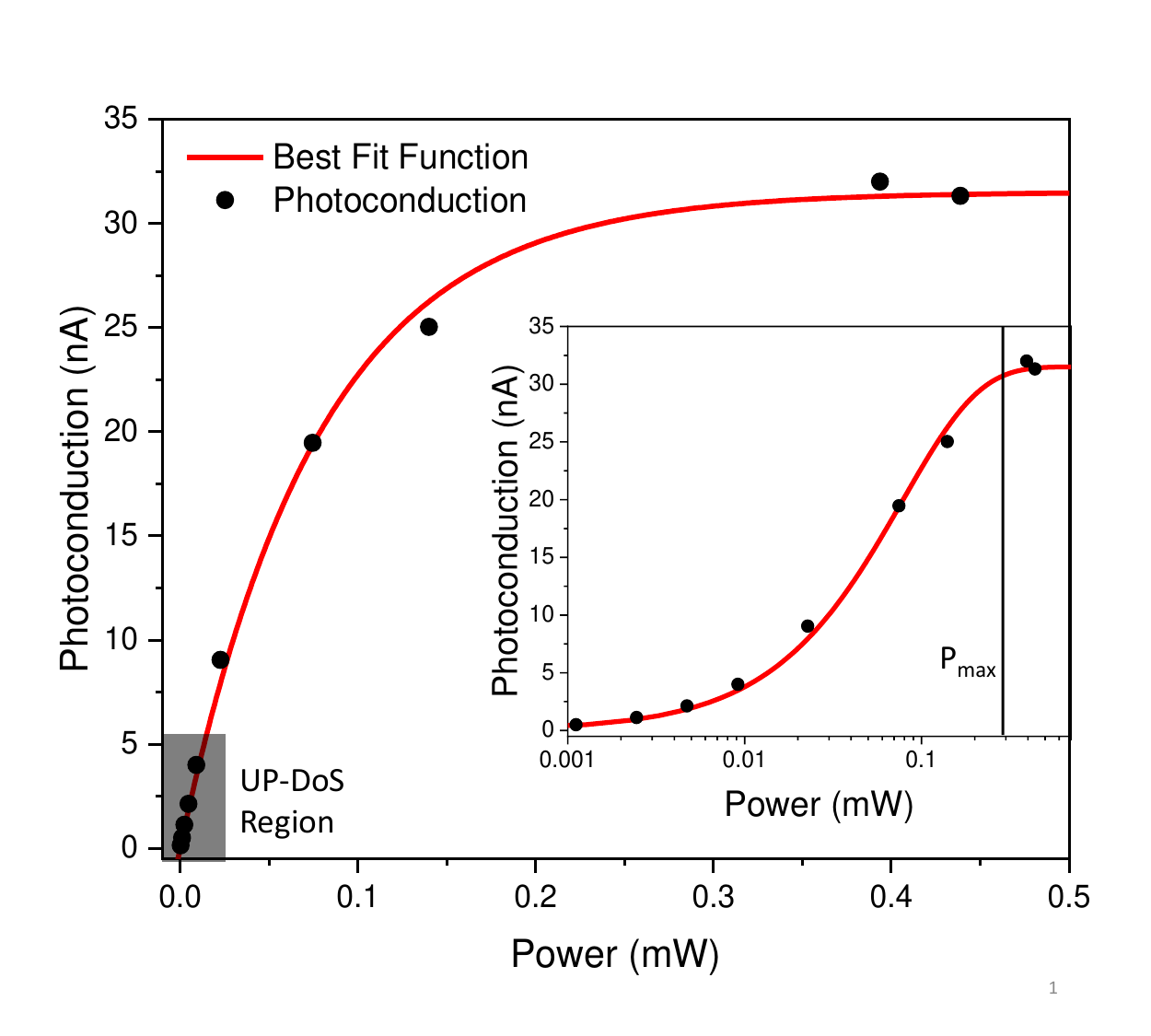}
   \end{tabular}
   \end{center}
   \caption[example] 
   {\textbf{Saturating Photocondcution} The photoconduction (PC) (\textit{black}) as a function of power measured on a copper oxide TFT with E$_{h\nu}$ = 1.4 eV, along with an exponential best-fit function (\textit{red}). UP-DoS is measured where the PC has a quasi-linear relationship with power. P$_{max}$ is found when the PC signal starts to saturate and no meaningful change in photoconduction occurs with further increase in power. } 
   \end{figure} 
\newpage
\section{SnO TFT Processing Conditions And DoS}
As seen in Fig. 4, device labeled p-type (black) differ in processing compared to purple and blue by four conditions: (1). p-type (black) TFT has 30 nm SnO instead of 20 nm deposition, (2). 100 $^{\circ}$C Sn precursor temperature instead of 120 $^{\circ}$C, (3). 250$^{\circ}$ C anneal temperature compared to 350 $^{\circ}$C, and (4) No post-TFT anneal, compared to TFT anneal 350$^{\circ} $C (purple) and 250 $^{\circ}$C (blue). See Table S.I for a full summary of device processing, where D1, E1, and E2 correspond to black, blue, and purple TFTs shown in Fig \ref{fig:p-type}a, respectively. Additionally, a second depletion mode device, D2, is shown with a DoS comparison in supplemental Fig. S2 with its processing conditions also seen in Table S.I.
 \begin{figure} [hbt]
   \begin{center}
   \begin{tabular}{c}
   \includegraphics[height=10cm]{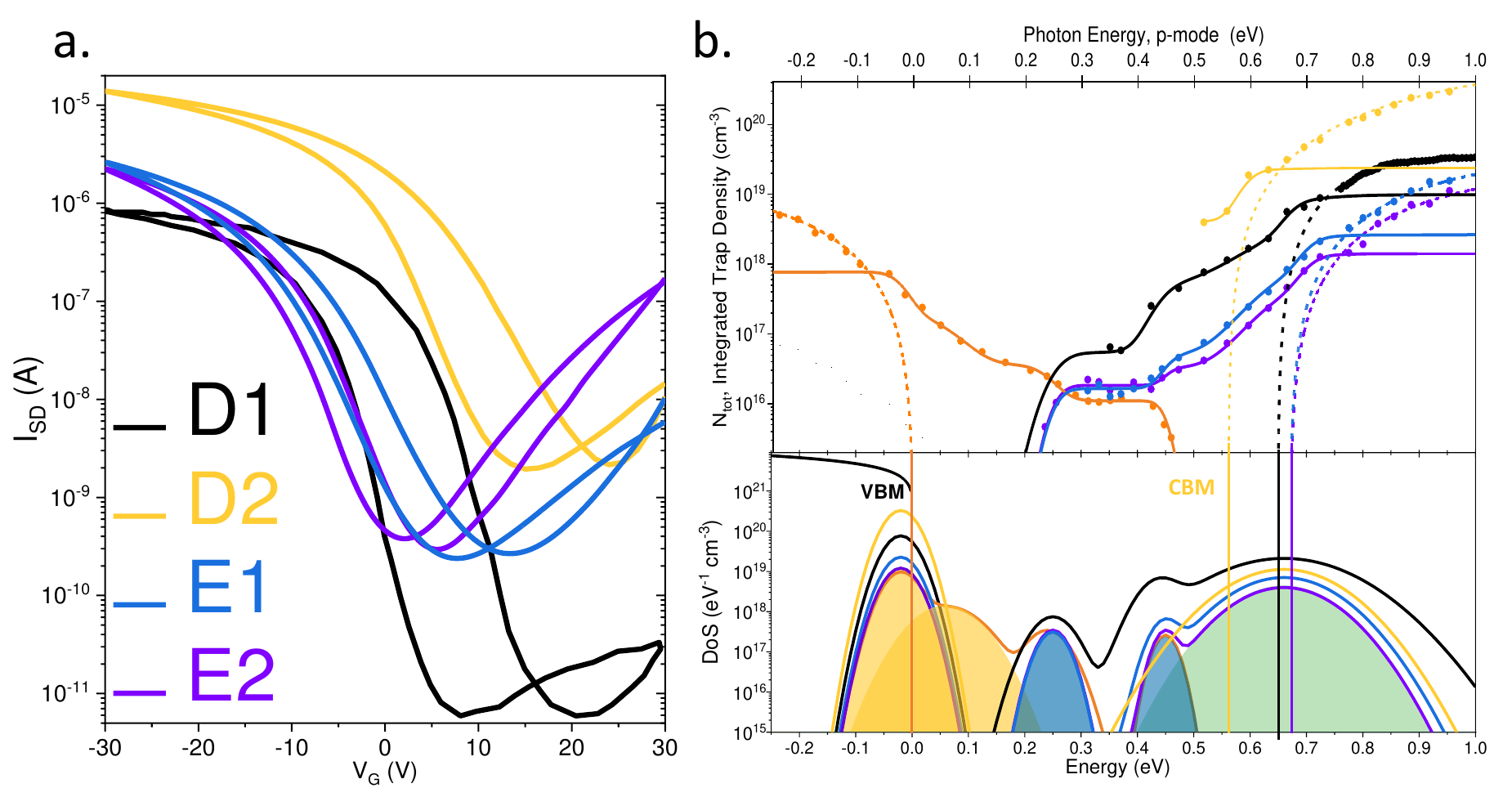}
   \end{tabular}
   \end{center}
   \caption[example] 
   {\textbf{SnO Subgap DoS} \textbf{(a)} Transfer curves for SnO TFTs listed in table S1. \textbf{(b)} The integrated trap density, N$_{tot}$, and the corresponding subgap DoS for the 4 SnO TFTs. D2 shows the most positive threshold among the 4 samples attributed to the large V$_{Sn}$+H DoS measuered. Additionally, D2 exhibits an exceptionally low bandgap of only $0.58$eV which may explain the high off current.}
   \end{figure} 
   
   \begin{center}
\begin{table*}[hbt]
 \begin{tabular}{c |c| c | c|c|c|c|c|c} 
Device & SnO thickness & SnO Precursor Temp. & SnO Annealing Temp. & TFT annealing temp.  & V$_{th}$  & Avg. $\mu_{FE}$ & ON-OFF & S.S \\
 (color) & (nm) & ($^{\circ}$ C) & ($^{\circ}$ C) & ($^{\circ}$C)  & (V) & (V cm$^{-1}s^{-1}$) & & (V/dec)\\

 \hline

D1 (black) &  30 & 100 & 250 & No annealing & 1 & 0.2 & $\sim 10^5$ & 2.6\\

D2 (yellow) & 30 & 100 & 350 & No annealing & 8 & 2.3 & $\sim 10^4$ & 4.6\\

E1 (blue) & 20 & 120 & 350 & 250 & -10 & 0.8 & $\sim 10^4$ & 5 \\

E2 (purple)& 20 & 120 & 350 & 350 & -13 & 0.8 & $\sim 10^4$ & 5 \\



 \end{tabular}
  \caption[example]
 { \label{allsno} 
The processing and growth conditions, as well as the TFT characteristics, for all SnO TFTs in this study.
}
 \end{table*}
\end{center}
\newpage
\section{Ambipolar SnO Fall Times}
\begin{figure}[H]
    \centering
    \includegraphics[height=6cm]{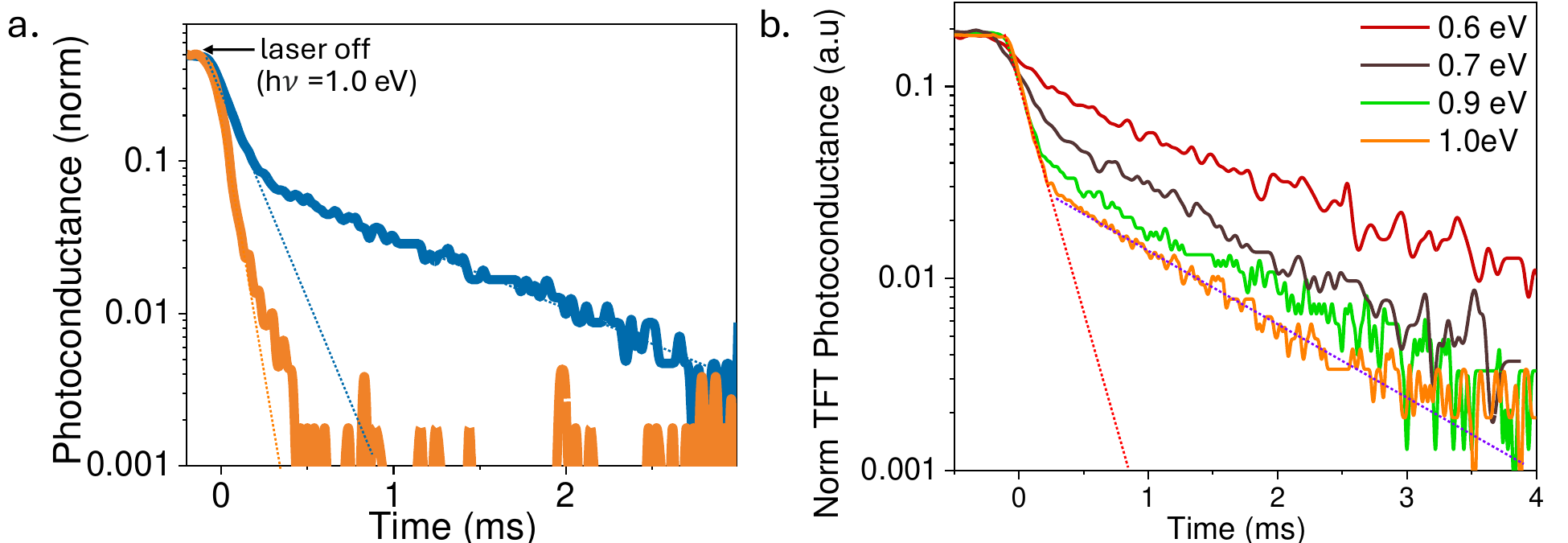}
    \caption{\textbf{Ambipolar SnO Falling Kinetics} \textbf{a.} Above gap excitation at 1.0 eV  photoconduction kinetic trace in n-mode (orange) and p-mode (blue). \textbf{b.} Ambipolar SnO TFT photoconduction kinetic trace from a $h\nu$= 1 eV to $h\nu$= 0.6 eV laser illumination that is modulated off at t = 0 to extract fall-times. The total fall time increases as a function of decreasing energy due to the removal of the much quicker above-gap component (red dashed) so that defect fall times dominate the signal (purple dashed).
    } 
    \label{fig:lifetime}
\end{figure}

\bibliography{mybib.bib}